\documentclass[a4paper,fleqn,usenatbib]{mnras}
\usepackage{graphicx}	
\usepackage{amsmath}	
\usepackage{amssymb}	
\usepackage[T1]{fontenc}
\usepackage{ae,aecompl}
\usepackage{graphicx}
\usepackage{newtxtext,newtxmath}
\usepackage[usenames]{color}
\usepackage{color}

\newcommand{\famA}{\mathcal{A}}
\newcommand{\famB}{\overline{\mathcal{A}}}
\newcommand{\famC}{\mathcal{P}}
\newcommand{\famD}{\overline{\mathcal{P}}}
\newcommand{\famO}{\mathcal{O}}

\title[Hierarchical Stationary Points]{The Stationary Points of the Hierarchical Three Body Problem}
\author[Hansen \& Naoz]{
	Bradley M. S. Hansen,$^{1,2}$\thanks{E-mail:hansen@astro.ucla.edu}, 
	\& Smadar Naoz$^{1,2}$
	\\	
	$^{1}$Mani L. Bhaumik Institute for Theoretical Physics, Department of Physics and Astronomy, University of California, Los Angeles, CA, 90095, USA\\
$^{2}$Department of Physics and Astronomy, University of California, Los Angeles, CA 90095, USA \\
}

\date{Accepted XXX. Received YYY; in original form ZZZ}

\pubyear{2020}

\begin{document}
	
	
	\maketitle

	\begin{abstract}	
	We study the stationary points of the hierarchical three body problem in the planetary limit ($m_2, m_3 \ll m_1$) at both the quadrupole and octupole orders. We demonstrate that the extension to octupole order preserves the principal stationary points of the quadrupole solution in the limit of small outer eccentricity $e_2$ but that new families of  stable fixed points occur in both prograde and retrograde cases. 
	The most important new equilibria are those that branch off from the quadrupolar solutions and extend to large $e_2$. The apsidal alignment of these families is a  function of mass and inner planet eccentricity, and is determined by the relative directions of precession of $\omega_1$ and $\omega_2$ at the quadrupole level.
	These new equilibria are also the most resilient to the destabilizing effects of relativistic precession.
	We find additional equilibria that enable libration of the inner planet argument of pericentre in the limit of radial orbits and recover the non-linear analogue of the Laplace-Lagrange solutions in the coplanar limit.
Finally, we show that the chaotic diffusion and orbital flips identified with the Eccentric Kozai Lidov mechanism and its variants can be understood in terms of the stationary points discussed here.

	
	\end{abstract}

	
	\section{Introduction}

        The application of the hierarchical three-body problem to planetary systems has received significant
attention over the last decade, motivated by the possibility that planets discovered in short period orbits \citep{MQ95,DJ18}
 may be the result of high eccentricities generated by special classes of solutions 
of the hierarchical problem \citep{IZ97,WM03,FT07,WM07,VF10,Naoz11,Naoz+12,NF13,Naoz16}.

A related question is the source of the high eccentricities observed in exoplanet systems in general \citep{MBV99,MBF05,US07}. Whatever the
original mechanism for eccentricity excitation is, only a subset of planetary systems are expected to undergo the
extreme eccentricity growth that would enable the tidal drag-down of planets to short periods.  The remainder of the population is expected to remain in a configuration set in place by the original eccentricity excitation. In some cases, this may be reflected in mean motion resonances, but many systems will exhibit oscillations in eccentricity and inclination modulated by the secular interactions between the planetary orbits.
	

The exoplanetary systems  are  complex dynamical systems that can exhibit a range of phenomena, including apsidal and
nodal circulation and libration and resonances between various periodicities in the system. One way to organise this information is to study the stationary points of the dynamical system and to classify the resulting equilibria in terms of their stability.
 Several studies have sought to understand how observed planetary systems fall within the range of available equilibria \citep{MFB06,LH06,LH07,LT09,VRL19}. We wish to examine the full range of stationary points relevant to the hierarchical secular problem, over the full range of mass ratio and mutual inclinations, extending extant results such as those by \cite{MG09} (limited to mass ratios of order unity) and \cite{MG11} (limited to prograde mutual inclinations). Our goal is to better understand the relationship between observed systems and the dynamical pathways by which systems can evolve.

 In \S~\ref{sec:Quad} we will review prior work on the stationary points of the hierarchical problem at quadrupolar order, and in \S~\ref{sec:Octo} we will extend this to the Octopolar order. We will examine the effect of corrections due to relativistic precession in \S~\ref{sec:Relative} and classify the stability of the identified stationary points in \S~\ref{sec:Stability}. In \S~\ref{sec:discussion} we will frame several well-known features of the dynamics in terms of the identified stationary points and summarise the conclusions in \S~\ref{sec:Conclude}.
	
	\section{Quadrupolar Fixed Points}\label{model}
\label{sec:Quad}	

In order to establish our framework, we review here the stationary points of the hierarchical three body
problem as described by the Hamiltonian expanded to quadrupole order in $\alpha_{12} = a_1/a_2$, the ratio
of the semi-major axes of the inner and outer planets \citep{Kozai62,Har68,LZ76,FO94}. Adopting the formulation
from \citet{Naoz16}, the equations of motion for the inner planet eccentricity, $e_1$, and the inner argument
of perihelion, $\omega_1$ are
	\begin{eqnarray}
\dot{e}_1 & = & 30 C \sqrt{1-e_1^2} \left( 1 - \theta^2 \right) e_1 \sin 2 \omega_1  \label{eqn:quadedot}\\
\dot{\omega}_1 & = & \frac{6 C }{\sqrt{1 - e_1^2}} \left[ 4 \theta^2 + \left(1 - \theta^2 - e_1^2 \right)
\left(5 \cos 2 \omega_1 - 1\right) \right. \nonumber \\
&& \left. + \mu_{12} \alpha_{12}^{1/2} \frac{\sqrt{1 - e_1^2}}{\sqrt{1 - e_2^2}} \theta \left( 2 + 3 e_1^2 - 5 e_1^2 \cos 2 \omega_1 \right) \right] \label{eqn:quadwdot}
	\end{eqnarray}
where $C$ is a constant that depends on the masses and semi-major axes and $\theta = \cos i_{\rm tot}$, where
$i_{\rm tot}$ is the relative inclination of the inner and outer orbital planes. The constant $\mu_{12}=m_1/m_2$
is the mass ratio between the inner and outer planet. This is the limiting expression in the planetary case,
where $m_1$ and $m_2$ are much less massive than the central body $m_0$. In the more general case, $\mu_{12}$
contains a prefactor $ m_0 (m_0 + m_1 + m_2)^{1/2}/(m_0+m_1)^{3/2}$.
	
The stationary points of the problem are found by solving for $\dot{e}_1=0$ and $\dot{\omega}_1=0$ simultaneously.
The fact that the argument of periastron of the outer planet -- $\omega_2$-- does not appear here (the `happy coincidence' of \citet{LZ76}) means that these two criteria are sufficient. A finite value of $e_2$ will not affect the dynamics but can parameterise the solutions.

Examination of equation~(\ref{eqn:quadedot}) establishes several possible branches of solution.
The most obvious are for $\omega_1=0$ (we will call this case $Q_1$) and $\omega_1=\pi/2$ (we will call this 
case $Q_2$). However, there are also a set of limiting cases that may also apply. This equation is also
satisfied in the case of circular ($e_1=0$ -- case $Q_{\rm C}$), radial ($e_1=1$ -- case $Q_{\rm R}$) or coplanar ($\theta=\pm 1$ -- case $Q_{||}$).
Let us discuss each in turn (Table~\ref{StableTableQ} presents a summary). 

\subsection{Case $Q_1$: $\omega_1=0$}
\label{sec:Q1}

By setting $\omega_1=0$ in equation~(\ref{eqn:quadwdot}), the condition for a stationary point is 
\begin{equation}
 \theta = - \frac{2}{\mu_{12} \alpha^{1/2}_{12}}\frac{\sqrt{1 - e_2^2}}{\sqrt{1 - e_1^2}} \ . \label{eqn:Quad1}
\end{equation}
We see that this family of stationary points only applies for retrograde orbits ($\theta<0$ always) and only
for large enough mass ratios, since $\theta>-1$ implies $\mu_{12} \alpha_{12}^{1/2} > 2$ for $e_1=e_2=0$, which is
when the family of stationary points first manifests itself. 

\subsection{Case $Q_2$: $\omega_1=\pi/2$}

In this case, setting $\dot{\omega}_1=0$ yields a quadratic solution for $\theta$ in terms of $e_1$.
\begin{equation}
\theta = \frac{\mu_{12} \alpha_{12}^{1/2}}{10} (1 + 4 e_1^2) \frac{\sqrt{1-e_1^2}}{\sqrt{1-e_2^2}}
\left[ - 1 \mp \left( 1 + \frac{60 (1 - e_2^2)}{\mu_{12}^2 \alpha_{12} ( 1+ 4 e_1^2)^2} \right)^{1/2}\right] \ . \label{eqn:Quad2}
\end{equation}
This condition defines the stationary point corresponding to the well-known Kozai-Lidov librations \citep{Kozai62,Lidov62}.
The low mass ($\mu_{12} \rightarrow 0$) limit asymptotes to the solution from the original (inner test particle) 
formulation by Kozai \& Lidov,
\begin{equation}
\theta = \pm \sqrt{\frac{3}{5}} \left( 1 - e_1^2 \right)^{1/2} \ . \label{eqn:Kozai}
\end{equation}
In the opposite limit of large $\mu_{12}$, this family of stationary points becomes
 asymmetric and the prograde and retrograde branches have different asymptotes, namely
\begin{equation}
\theta \sim  \frac{3}{\mu_{12} \alpha_{12}^{1/2}} \frac{\sqrt{1-e_1^2}}{1 + 4 e_1^2} \left( 1 - e_2^2 \right) \quad ({\rm{ prograde}}) \ , \label{eqn:Quad2_pro}
\end{equation}
which tends to polar orbits (regardless of $e_1$) in the limit of an outer test particle \citep{Zig75}, and 
\begin{equation}
\theta \sim - \frac{1}{5} \mu_{12} \alpha_{12}^{1/2} ( 1 + 4 e_1^2) \frac{\sqrt{1 - e_1^2}}{\sqrt{1 - e_2^2}} \quad ({\rm{ retrograde}})  \ , \label{eqn:Quad2_ret}
\end{equation}
which becomes unphysical  (because $\mu_{12}$ is large) unless $e_1$ is close enough to radial.
 Thus, the retrograde stationary points only exist for almost radial orbits in the large mass ratio limit.

Given this asymmetry at high mass ratios, we will adopt separate labels for the prograde ($Q_2^+$) and
retrograde ($Q_2^-$) parts of the stationary point family.

\begin{table}
\centering
\begin{minipage}{140mm}
\caption{Classification of the Quadrupolar Stationary Points.
 \label{StableTableQ}}
\begin{tabular}{@{}lccl@{}}
Name & Orbit & $\mu_{12} \alpha_{12}^{1/2}/\sqrt{1 - e_2^2} $ & Comments \\
\hline
$Q_1$ & Retrograde & $>2$ & Saddle Point \\
$Q_2^+$ & Prograde & all & Fixed Point \\
$Q_2^-$ & Retrograde & all  & Fixed Point\\
$Q_{\rm C}$ & Both  & all  & Saddle Point \\
$Q_{\rm R}$ & Polar & all & Saddle Point \\
$Q_{||}$ & Retrograde & $>1.8$ & Saddle Point\\
\hline
\end{tabular}
\end{minipage}
\end{table}

\subsection{Case $Q_{\rm C}$: $e_1=0$}
\label{sec:Qc}

An alternative path to satisfy $\dot{e}_1=0$ is to set $e_1=0$ in 
equation~(\ref{eqn:quadedot}). In this case, it does not impose a condition 
on $\omega_1$, as in the previous two sections. Instead, we must constrain $\omega_1$ by setting equation~(\ref{eqn:quadwdot}) to zero,
which yields
\begin{equation}
\cos 2 \omega_1 = \frac{1}{5} \left[ 1 - \frac{2 \theta}{1 - \theta^2} \left( 2 \theta + \frac{\mu_{12} \alpha_{12}^{1/2}}{\sqrt{1 - e_2^2}}
\right) \right] \ . \label{eqn:QuadC}
\end{equation}
We will discuss the meaning of this equation in \S~\ref{sec:QuadStable}, but it is worth noting here that setting $\cos 2 \omega_1=1$
yields the same equation as the circular limit of equation~(\ref{eqn:Quad1}), and setting $\cos 2 \omega_1=-1$ yields
the same equation as the circular limit of equation~(\ref{eqn:Quad2}).

\subsection{Case $Q_{\rm R}$: $e_1=1$}
\label{sec:QR}

To satisfy $\dot{\omega}_1=0$ for $e_1=1$ we require both $\theta=0$ and $\omega_1=0$ or $\pi$. This case is
therefore a very localised stationary point -- in the limit of polar, radial orbits.

\subsection{Case $Q_{||}$: $\theta=\pm 1$}
\label{sec:Qplanar}

Setting $\dot{e}_1=0$ and $\dot{\omega}_1=0$ in Eqs.~(\ref{eqn:quadedot}) and (\ref{eqn:quadwdot}), in
the prograde, coplanar ($\theta=1$) case, we do not find any 
any physical solutions.
For the case $\theta=-1$, these equations yield the
condition
\begin{equation}
\cos 2 \omega_1 = \frac{1}{5 e_1^2} \frac{ 4  + e_1^2 - \mu_{12} \alpha_{12}^{1/2} \sqrt{1 - e_1^2} ( 2 + 3 e_1^2)/\sqrt{1 - e_2^2}}
{1 - \mu_{12} \alpha_{12}^{1/2} \sqrt{1 - e_1^2}/\sqrt{1 - e_2^2}} \ . \label{eqn:quadcop}
\end{equation}
This equation yields physically realistic solutions ($| \cos 2 \omega_1 |<1$) for $\mu_{12} \alpha_{12}^{1/2} > 1.8$
and has the same $e_1 \rightarrow 0$ limit as Case $Q_1$, suggesting a common link between these cases.
	
\subsection{Precession of the Outer Body}

Although the evolution of the system, to quadrupolar order, does not depend on $\omega_2$, this variable will come into play when we extend our analysis to octupole order. Therefore, the condition $\dot{\omega}_2=0$, at quadrupolar order, will become relevant. In the case of $\omega_1=\pi/2$, this leads to the condition
\begin{equation}
    2\theta \left(1 + 4 e_1^2\right) + \mu_{12} \alpha_{12}^{1/2} \sqrt{\frac{1-e_1^2}{1-e_2^2}} \left[
    2 + 3 e_1^2 + \left( 5 \theta^2 - 3 \right) \left( 1 + 4 e_1^2 \right) \right] = 0 \ ,
 \end{equation}
which has the solution
\begin{equation}
    \theta = \frac{1}{5 \mu_{12} \alpha_{12}^{1/2}} \sqrt{\frac{1-e_2^2}{1-e_1^2}} 
    \left[ -1 \pm \left( 1 + 5 \mu_{12}^2 \alpha_{12} \frac{1-e_1^2}{1-e_2^2}
    \frac{1+9 e_1^2}{1+ 4 e_1^2}\right)^{1/2}\right] \ . \label{eqn:Quadouter}
\end{equation}
This condition becomes relevant in the limit of large $\mu_{12}$, so the asymptotic solution is  
\begin{equation}
 \theta^2 = \frac{1}{5} \left( \frac{1 + 9 e_1^2}{1+4 e_1^2} \right) . 
\end{equation}
The equivalent solution in the $\omega_1=0$ case is
\begin{equation}
     \theta = \frac{1}{5 \mu_{12} \alpha_{12}^{1/2}} \sqrt{\frac{1-e_2^2}{1-e_1^2}} 
    \left[ -1 \pm \left( 1 + 25 \frac{\mu_{12}^2 \alpha_{12}}{1-e_2^2}
    \right)^{1/2}\right] \ . \label{eqn:Quadout2}
\end{equation}

\subsection{Nature of the Stationary Points}
\label{sec:QuadStable}

\begin{figure}
                \centering
                \includegraphics[width=1.0\linewidth]{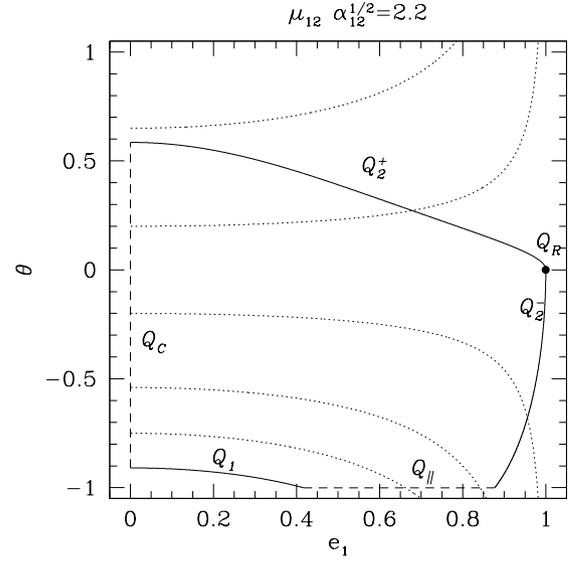}
                \caption{ The solid curves show the stationary point solutions $Q_1$, $Q_2^+$ and $Q_2^-$ for
the case $\mu_{12} \alpha_{12}^{1/2}=2.2$. The two dashed lines are the special case solutions $Q_{\rm C}$ and $Q_{||}$,
while the large solid point at $(e_1,\theta)=(1,0)$ is the special case $Q_{\rm R}$. The dotted lines indicate trajectories of constant $\theta \sqrt{1-e_1^2}$ -- the horizontal co-ordinate in Figure~B1. Each point in Figure~B1 is therefore labelled with any stationary points intersected by the corresponding dotted trajectory.
 } 
                \label{fig:Example}
        \end{figure}

Figure~\ref{fig:Example} shows how these different fixed point families are related to one another,
for the case  $\mu_{12} \alpha_{12}^{1/2}=2.2$. Stationary point families associated with a fixed $\omega_1$ ($Q_1$, $Q_2^+$ and $Q_2^-$) are shown as solid curves
while the families with a range of $\omega_1$ ($Q_{\rm C}$ and $Q_{||}$) are shown as dashed curves. The special case
family $Q_{\rm R}$ is shown as a solid point.
The stationary point family $Q_{\rm C}$ connects the prograde family $Q_2^+$ and the retrograde family $Q_1$.
The stationary point family $Q_{||}$ connects the retrograde families $Q_1$ and $Q_2^-$.


The stationary point family $Q_{\rm C}$ connects the prograde family $Q_2^+$ and the retrograde family $Q_1$.
The stationary point family $Q_{||}$ connects the retrograde families $Q_1$ and $Q_2^-$.
The various families shift as a function of mass and separation, and Figure~B1 of the online appendix shows a general overview
of these relationships, as a function of  $\mu_{12} \alpha_{12}^{1/2}$ and $x=\theta \sqrt{1 - e_1^2}$.

To understand the dynamics of the system near each of these points, we plot curves of constant energy, subject
to the constraint of angular momentum conservation. This latter condition establishes a relationship between
$\theta$ and $e_1$, such that 
\begin{equation}	
        G_0^2 = \mu_{12}^2 \alpha_{12} \left( 1 - e_1^2\right) + 2 \mu_{12} \alpha_{12}^{1/2} \sqrt{1 - e_1^2}\sqrt{1 - e_2^2} \, \theta \ .
\label{eqn:TotG}
\end{equation}

\begin{figure}
                \centering
                \includegraphics[width=1.0\linewidth]{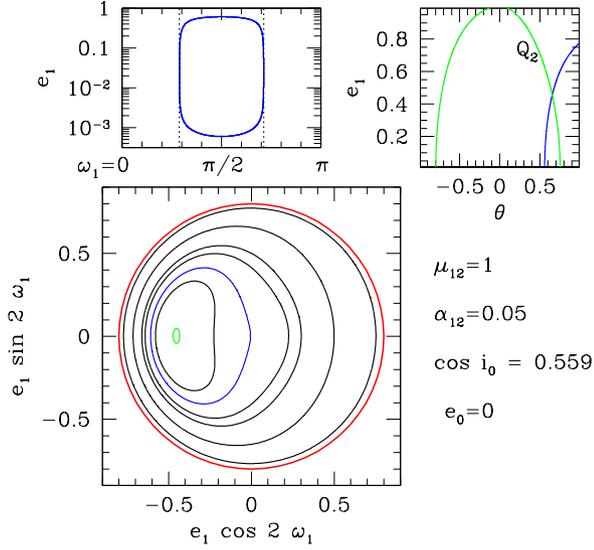}
                \caption{ The main panel shows curves of constant energy, subject to the constraint that the total angular momentum take a particular value. The blue curve has  the initial conditions given by $e_1=e_0$ and $i_{\rm tot}=i_0$. This shows a large libration about the fixed point at $\omega_1=\pi/2$. The panel to the upper right shows how $e_1$ and $\theta$ evolve along this trajectory (blue curve). The green curve in this diagram shows the $Q_2$ fixed point family (similarly, the green curve in the main panel shows the location of the fixed point). The panel in the upper left shows the result of an orbital integration (using the code of Naoz et al. (2013) ) for these parameters. The dotted lines indicate the $\omega_1$ obtained from equation~(\ref{eqn:QuadC}), and demonstrate that the stationary points $Q_{\rm C}$ represent a family of unstable saddle points.
                The red circle in the main panel indicates the maximum eccentricity that it is possible to achieve with this initial condition.
}
                \label{fig:mapq3}
        \end{figure}

Figure~\ref{fig:mapq3} shows the curves of $e_1(\omega_1)$ in the case of $\alpha_{12}=0.05$ and $\mu_{12}=1$. In
the panel in the upper right, the green curve shows the stationary point family $Q_2$ for these parameters. The blue
curve in the same diagram represents a curve of constant $G_0^2$, chosen such that $\theta=0.559$ for $e_0=0$. The
main panel then shows curves of constant energy, subject to the constraint that the angular momentum has the above
value. The blue contour is the one that corresponds to our chosen initial conditions, and the green contour illustrates the libration about the $Q_2^+$ family. This is the standard Kozai-Lidov family and therefore the $Q_2^+$ family is a stable equilibrium -- a `fixed point' family. This case represents the point $(0.559,0.224)$ in Figure~B1.

The panel in the upper left of Figure~\ref{fig:mapq3} also illustrates the nature of the $Q_{\rm C}$ family. The vertical 
dotted lines illustrate the value we get from equation~(\ref{eqn:QuadC}) for this case. Thus, the libration about
the $Q_a^2+$ point approaches $e_1 \sim 0$ along this value and then sweeps through $\omega_1$ until it emerges
at the other corresponding solution to the equation. This represents the change in angle of the blue contour as
it sweeps around the origin. Thus, family $Q_{\rm C}$ is {\it an unstable equilibrium} -- the saddle point at $e_1=0$. 

\begin{figure}
                \centering
                \includegraphics[width=1.0\linewidth]{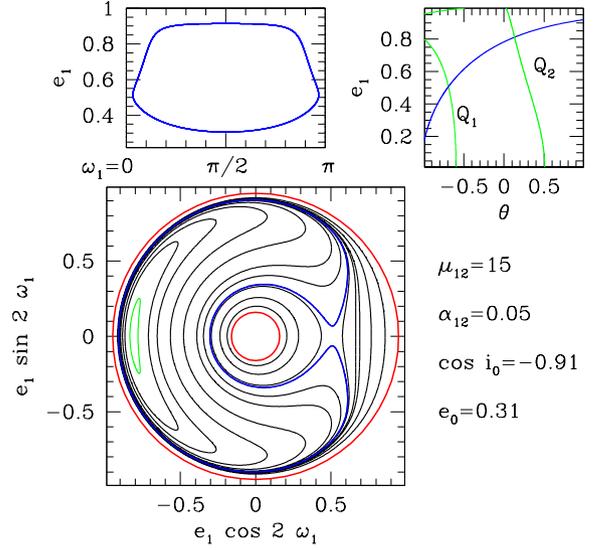}
                \caption{ The main panel shows curves of constant energy, subject to the constraint that the total angular momentum take a particular value. The blue contour shows the particular trajectory given by the initial conditions $e_1=e_0$
and $i_{\rm tot}=i_0$. The variation in $e_1$ and $\theta$ is shown by 
 the blue curve in the panel to the upper right. In this upper panel, there are now three green curves, which represent the relationships between $\theta$ and $e_1$ that corresponds to the fixed point family $Q_2$ and the stationary points family $Q_1$. The blue curve in the main panel again shows a large libration about the fixed point at $\omega_1=\pi/2$ as well as an avoidance of the saddle point at $\omega_1=0$ (the $Q_1$ family). The two red circles indicate the minimum and maximum eccentricities achievable with the given initial conditions.
The panel in the upper left shows the result of an orbital integration -- using the octupole code \citep{Naoz+13} -- for these parameters. The dotted lines indicate the $\omega_1 =0 $ and $\pi$, representing the location of $Q_1$. Note that this does not coincide with the minimum of $e_1$.  
}
                \label{fig:mapq2}
        \end{figure}

To understand the nature of the $Q_1$ family, we need to move to large $\mu_{12}$. Figure~\ref{fig:mapq2} shows the
case for $\alpha_{12}=0.05$ and $\mu_{12}=15$. We have chosen initial conditions here to provide a very large amplitude libration about the $Q_2^+$ fixed point, which actually approaches $\omega_1=0$. We see that this point (the location of the $Q_1$ family of stationary points) is a saddle point. Thus, $Q_1$ represents a family of unstable stationary points. 
This also results in a qualitative change in the nature of the dynamics. At lower mass ratios  (such as in Figure~\ref{fig:mapq3}), the choice of initial conditions implies either libration or circulation. For larger masses, the presence of the $Q_1$ family now divides the space into three parts -- the libration region encloses a region of inner circulation, with an outer circulation region at large $e_1$.

This figure  shows two red circles. The outer corresponds to the maximum eccentricity and the interior red circle in Figure~\ref{fig:mapq2} represents the minimum $e_1$ achievable with this angular momentum constraint, and occurs at $\theta=-1$ . These two limits can also be read off the blue curve in the upper right panel. The argument of periastron $\omega_1$ circulates in this case, so the minimum is not part of the $Q_{||}$ stationary point family.

 To understand the special family $Q_{||}$ we need to go to larger initial $e_1$. This is illustrated in 
 Figure~B2 of the online appendix, which shows the case for $\alpha_{12}=0.05$, $\mu_{12}=10$, $\theta=-1$ and $e_1=0.6$. There we demonstrate that the family $Q_{||}$ plays the same role as $Q_{\rm C}$, in the high mass ratio limit where the orbits never
get circular, but do approach the retrograde, coplanar limit.


So, we infer that the family $Q_{||}$ plays the same role as $Q_{\rm C}$, in the high mass ratio limit where the orbits never 
get circular, but do approach the retrograde, coplanar limit.

The only special case left is the point $Q_{\rm R}$. Integrations that start close to $e_1 =1$ and $\theta=0$ avoid the
limit, oscillating to $e_1 \sim 1$ but $\theta \sim \pm 1$, so O$_{\rm R}$ is a saddle point. This is despite the fact 
that the stationary point intersects the continuation of the prograde and retrograde branches of the $Q_2$ family.
This saddle point is therefore the ultimate cause for why one does not get flips of the orbital plane at the
quadrupolar level of approximation.

The classification of the stability of the various quadrupolar stationary points (whether solutions librate about the
equilibrium -- a fixed point -- or avoid the equilibrium location -- a saddle point) are summarized in Table~\ref{StableTableQ}.

 \section{Stationary Points at the Octopolar Level}
\label{sec:Octo}

The neatness of the quadrupolar analysis relies, in part, on the fact that $\omega_2$ does not appear in the Hamiltonian, and so we need only satisfy $\dot{e}_1=0$ and $\dot{\omega}_1=0$. This is no longer true when the expansion is taken to octupole order, and so we must now consider, in addition, $\dot{e}_2=0$ and $\dot{\omega}_2=0$. However, as noted by several authors \citep[e.g.,][]{NF13,LN14,Antognini15,Naoz16}, the timescale for changes in $e_2$ and $\omega_2$ are usually much longer than for $e_1$ and $\omega_1$, so that the short-term dynamics is often regulated by a stationary point of only $\dot{e}_1=0$
and $\dot{\omega}_1=0$. This has the consequence that the stationary point families of the octupolar problem are intimately related to those of the quadrupolar problem.

Thus, we are searching for stationary points of the system given by equations~(77), (78), (73) and (74) of \cite{Naoz16}, reprinted in appendix~\ref{eqn:octupole}, in the planetary limit. As in \S~\ref{sec:Quad}, the stationary points can be classified in terms of particular values of $\omega_1$ and, now also, $\omega_2$. We must also consider the quantity $\epsilon = \alpha_{12} e_2/(1-e_2^2)$ when classifying these equilibria. This parameter quantifies the strength of the octupole term, and we will adopt $\epsilon<0.1$ as the criterion for
restricting our analysis to the octupole level.
A  larger value of $\epsilon$ would  require extending the expansion to higher orders to achieve accuracy. \citep[e.g.,][]{Hamers+16,Will17} 

 We will also require a naming convention to conveniently identify particular stationary point families. Those families associated with $\omega_1=0$ and $\omega_2=0$ will be designated as  $\mathcal{A}$ -- because $\omega_1$ and $\Omega_1$ are aligned. The apsidally anti-aligned case ($\omega_2=\pi$) will be called $\famB$. 
 Those families with $\omega_1=\pi/2$ will be designed as  $\famC$ ($\omega_2=\pi/s$)
and $\famD$ ($\omega_2=3 \pi/s$) respectively (because $\omega_1$ and $\Omega_1$ are perpendicular in this case).

\begin{table*}
\centering
\begin{minipage}{140mm}
\caption{Classification of Octupole Level Stationary Points. The quantity $\beta=\mu_{12} \alpha_{12}^{1/2}/\sqrt{1-e_2^2}$.
We define critical inclination $I_{\rm crit}$ as the inclination of a solution in the $e_1 \rightarrow 0$ limit.
Family $\famA$ represents $(\omega_1,\omega_2)=(0.0)$, while $\famB$ represents $(0,\pi)$, $\famC$ represents $(\pi/2,\pi/2)$
and $\famD$ represents the case $(\pi/2,3\pi/2)$.
 \label{StableTable}}
\begin{tabular}{@{}llll@{}}
Name & Orbit &  Stability & Comments \\
\hline
$\famA_1$ & Prograde &  Saddle point & Critical inclination starts at $I_{\rm crit}=90^{\circ}$ at low masses 
 and tends to $I_{\rm crit}=63.75^{\circ}$ at high $\mu_{12}$ \\
$\famA_2$ & Retrograde & Saddle point & Appears for $\beta>4$, with $I_{\rm crit}=120^{\circ}$ initially, but asymptotes 
 to $I_{\rm crit}=116.25^{\circ}$ at large $\mu_{12}$  \\
$\famB_{\rm Q}$ & Retrograde &  Saddle point & Generalisation of quadrupolar family $Q_1$. Requires
 $\beta>2.236$. 
  Critical inclination starts \\
& & & at $I_{\rm crit}=180^{\circ}$ and evolves to $I_{\rm crit}=90^{\circ}$ in
high mass limit.\\
$\famB_1$ & Retrograde & Saddle point & $I_{\rm crit}=180^{\circ}$ for $\beta=0.5$ and decreases to $120^{\circ}$ at $\beta=3.35$, 
 where it is subsumed by \\
& & & $\famB_{\rm Q}$ in the circular limit. \\
$\famC_{\rm Q}^+$ & Prograde & Fixed point &  Generalisation of Quadrupolar family Q$_2^+$. Critical inclination
is 39.23$^{\circ}$ for low $\mu_{12}$, \\
& & & increasing to 65.75$^{\circ}$ at $\beta=5.25$, where it switches apsidal alignment. \\
$\famC_{\rm Q}^-$ & Retrograde & Fixed point & Generalisation of Quadrupolar family $Q_2^-$. Critical inclination
is 140.76$^{\circ}$ for low $\mu_{12}$, \\
& & & increasing to 180$^{\circ}$ at $\beta=2.254$. For larger $\beta$ it does not reach $e_1=0$. \\
$\famC_1$ & Prograde & Fixed point & Most robust family in face of relativistic precesson. $I_{\rm crit} \rightarrow 
63.43^{\circ}$ in the high mass limit. \\
$\famC_2$ & Retrograde & Fixed point & Appears when $\beta=0.773$, with $I_{\rm crit}=148.9^{\circ}$. Tracks $\famC_Q$$^-$ but
extends to large $e_1$ \\
$\famC_3$ & Retrograde &  Fixed point & Satisfies $\epsilon<0.1$ for $\beta>0.314$. For $\beta>0.46$ this has a solution at  $e_1=0$, with  \\
& & &critical angle $I_{\rm crit}=180^{\circ}$ \\
 & Retrograde &  Saddle point & For $\beta>0.653$, this becomes a saddle point, at $I_{\rm crit}=148.9^{\circ}$.  
The critical angle decreases \\ & & &  with increasing mass, tending to $I_{\rm crit}=116.56^{\circ}$ in the high mass limit. \\ 
$\famC_{\rm R}$ & Prograde & Inner Fixed point & At low masses, $80.7^{\circ}<I<90^{\circ}$ for $\epsilon<0.1$ and narrows
for $\mu_{12}>2$.\\
$\famD_{\rm Q}$$^+$ & Prograde & Fixed point & Generalisation of Quadrupolar family $Q_2^+$, which
reaches $e_1=0$ for $\beta>5.25$ and \\
& & &  $I_{\rm crit}=65.75^{\circ}$. In the high mass limit, $I_{\rm crit} \rightarrow 90^{\circ}$. \\
$\famD_{\rm Q}$$^-$ & Retrograde & Fixed point & Quadrupole and Octupole apsidal switches limit to $0.794 < \beta < 1.738$ \\
$\famD_1$ & Prograde & Fixed point & Emerges for $\beta>0.47$ for $\epsilon<0.1$.\\
$\famD_2$ & Retrograde &  Fixed point & Appears when $\beta=0.773$, with $I_{\rm crit}=148.9^{\circ}$. Tracks $\famC_Q$$^-$ but
extends to large $e_1$ \\
$\famD_3$ & Prograde &  Saddle point & $I_{\rm crit} \sim 90^{\circ}$ for low $\mu_{12}$, but drops to 65.75$^{\circ}$
at $\beta=5.25$. \\
$\famD_4$ & Retrograde &   Fixed Point & Large $e_1$ but not as large as $\famD_{\rm R}$. Appears for $\beta>0.492$, assuming $\epsilon<0.1$.\\
$\famD_{\rm R}$ & Retrograde & Inner Fixed point & At low masses, $90^{\circ}<I<99.3^{\circ}$ for $\epsilon<0.1$ and narrows
for $\mu_{12}>2$. \\
$\famO_{\rm C}$ & Prograde &  Saddle point & This is related to the $I_{\rm crit}$ limits of other families noted above. \\
$\famO_{\rm LL}$ & Prograde & Fixed point & This is generalisation of  the Laplace--Lagrange solutions. \\
$\famO_{||}$ & Retrograde & Saddle point & Generalisation of $Q_{||}$ which it closely resembles. \\
$\famO_{\rm R}$ & Polar &  Saddle point & In the coplanar limit, the solution is localised unless $e_2$ exceeds a threshold. \\
\hline
\end{tabular}
\end{minipage}
\end{table*}

\subsection{Case $\mathcal{A}$: $\omega_1 = \omega_2=0$}
\label{sec:CaseA}

If we set $\omega_2=0$, then $\dot{e}_1=0$ and $\dot{e}_2=0$ if $\omega_1=0$. The same conditions apply if
$\omega_2=\omega_1=\pi$. In this instance, the conditions $\dot{\omega}_1=0$ and $\dot{\omega}_2=0$ amount to 
\begin{equation}
 2 + \beta \theta  =  
\pm \frac{25}{8}\frac{\alpha_{12}\theta e_1 e_2}{1-e_2^2} 
\left[ \frac{1 - 24 e_1^2 - 5 (1-3 e_1^2) \theta^2 }{10 \theta e_1^2}
-  
  \theta - \beta \right] \label{eqn:ifix1}
\end{equation}
and
\begin{eqnarray}
&& 2 \theta + \beta
\left[ 5  \theta^2  + \frac{6 e_1^2 -1}{1 - e_1^2} \right] =  
 \mp \frac{5}{8} \frac{\alpha_{12} e_1}{1-e_2^2}  \times \nonumber \\
&&\left[ 10 e_2 \theta 
\left( 1 + \beta \theta \right) 
 -  (1+4 e_2^2) \frac{\beta}{e_2}  
\left( \frac{ 1 -8 e_1^2}{(1-e_1^2)} - 5  \theta^2  \right)\right], \label{eqn:ifix2}
\end{eqnarray}
where $\beta = \mu_{12} \alpha_{12}^{1/2} \sqrt{(1-e_1^2)/(1-e_2^2)}$.
The upper sign on the right hand side of these two equations applies for this case (and the corresponding case where both angles are $\pi$). The lower sign
in these two equations applies for the case where $\omega_1=0$ and $\omega_2=\pi$ (or vice versa). This is the apsidally anti-aligned case and will be treated in the next section.

The numerical solution of equations~(\ref{eqn:ifix1}) and (\ref{eqn:ifix2}) yields two families of solution, which
are illustrated in Figure~B3 of the online appendix. The first is family $\famA_1$, which tends towards
polar orbits ($\theta \rightarrow 0$) for small $\mu_{12}$ but moves to larger $\theta$ as $\mu_{12}$ increasesi, asymptoting
to $\theta=1/\sqrt{5}$. The
$e_1$--$e_2$ relation is pretty steep and so this satisfies $\epsilon<0.1$ for only a limited range of $e_1$.

A second family of solutions emerges at large $\mu_{12}$, designated as $\mathcal{A}_2$ and also shown in Figure~B3.  In the limit of large $\mu_{12}$ this again trends towards $\theta \rightarrow 0$. In the high $\mu_{12}$ limit, we find that this family is restricted to a finite range of $e_1$.
This stationary point is clearly related to the dynamics identified by \cite{NLZ17} and \cite{DeL18} in the context of the outer test particle case. We will discuss this more in \S~\ref{sec:switch}.

\subsection{Case $\overline{\mathcal{A}}$: $\omega_1=0$, $\omega_2=\pi$}

\label{sec:CaseB}

This case represents the lower sign choice in equations~(\ref{eqn:ifix1}) and (\ref{eqn:ifix2}). In physical terms, it means that the arguments of periastron of the two planets are apsidally misaligned by 180$^{\circ}$. Examples of
 the resulting stationary point families are shown in Figure~B4 of the online appendix. 
Unlike for case~$\mathcal{A}$, viable solutions only start to appear for mass ratios $\mu_{12}>1.5$ (for $\alpha_{12}=0.05$). Formally, we can find stationary points at smaller mass ratios, but they all occur for $\epsilon>0.1$. This implies that such points may exist but a full description may require higher order terms. For large enough $\mu_{12}$, the solution extends to
low $e_1$ and $e_2$ and represents the complementary case for the $\famA_2$ solutions.


%

The first stationary point family to appear (as we increase $\mu_{12}$) is family $\overline{\mathcal{A}}_1$, which manifests in
 Figure~B4 as a family of retrograde orbits  and large $e_2$, and is restricted
to approximately circular orbits ($e_1 \ll 1$). As $\mu_{12}$ increases, this family moves towards more inclined
(but still retrograde) configurations, and with larger $e_1$. For large enough mass ratios, a second family appears,
which we term $\overline{\mathcal{A}}_{Q}$. This is because the properties of these stationary points bear a strong similarity to the quadrupole family $Q_1$ discussed in \S~\ref{sec:Q1}. The $e_1$--$\theta$ relation for $\overline{\mathcal{A}}_Q$ tracks almost exactly the equation~(\ref{eqn:Quad1}) for $Q_1$, for the relevant masses. The value of $e_2$, in this case, is small but not exactly zero.

As $\mu_{12}$ continues to increase, the two branches merge into a single continuous family, bounded from below
by a minimum $e_1$. For $\mu_{12}=50$,  the $\overline{\mathcal{A}}_1$ part of the curve is restricted to a
narrow range of $e_1$, in a similar fashion to family $\mathcal{A}_2$ of \S~\ref{sec:CaseA}.  The $\overline{\mathcal{A}}_{Q}$ curve also extends
to lower $e_1$, but there appears to be a gap. We will discuss this further in \S~\ref{sec:switch}.

\subsection{Case $\mathcal{P}$: $\omega_1 = \omega_2 = \pi/2$}

\label{sec:CaseC}

The third case occurs when both arguments of periastron are at right angles with respect to the line of nodes.
Once again, this choice of parameters automatically satisfies $\dot{e}_1=0$ and $\dot{e}_2=0$, leaving the following
conditions to locate the stationary points

\begin{eqnarray}
&& 5 \theta^2 - 3 (1 - e_1^2) + \beta ( 1+ 4 e_1^2) \theta =
\mp \frac{5 \alpha_{12} e_2}{16 ( 1 - e_2^2)} \times \nonumber \\
&& \left[ e_1 \left( \theta + \beta \right) 
 \left( 15 (3 + 4 e_1^2) \theta^2 - 11 - 17 e_1^2 \right) - \right. \nonumber \\
&& \left. \frac{(1-e_1^2)}{e_1} \theta
\left( 11 + 51 e_1^2 - 15 (1 + 4 e_1^2) \theta^2 \right) \right], \label{eqn:ifix3}
\end{eqnarray}

and

\begin{eqnarray}
&& 2 \theta + \beta
\left(  5 \theta^2  - 3 + \frac{2+3 e_1^2}{1+4 e_1^2} \right) = \pm \frac{5}{8} \frac{\alpha_{12} e_1}{(1-e_2^2) (1+ 4 e_1^2)} \times \nonumber \\
&&\left[ \frac{(1+4e_2^2)}{e_2} \beta  \theta \left( 11 + 17 e_1^2 - 5 (3 + 4 e_1^2) \theta^2 \right) \right. \nonumber \\
&&\left. + e_2 \left( 1 + \beta \theta \right) \left( 11 + 17 e_1^2 - 15 (3 + 4 e_1^2) \theta^2  \right) \right].
\label{eqn:ifix4}
\end{eqnarray}

As in the previous sections, the upper sign in equations~(\ref{eqn:ifix3}) and (\ref{eqn:ifix4}) refers to the case
of apsidal alignment, while the lower sign represents the anti-aligned case ($\omega_1=3 \pi/2$, $\omega_2=\pi/2$).

\begin{figure}
                \centering
                \includegraphics[width=1.0\linewidth]{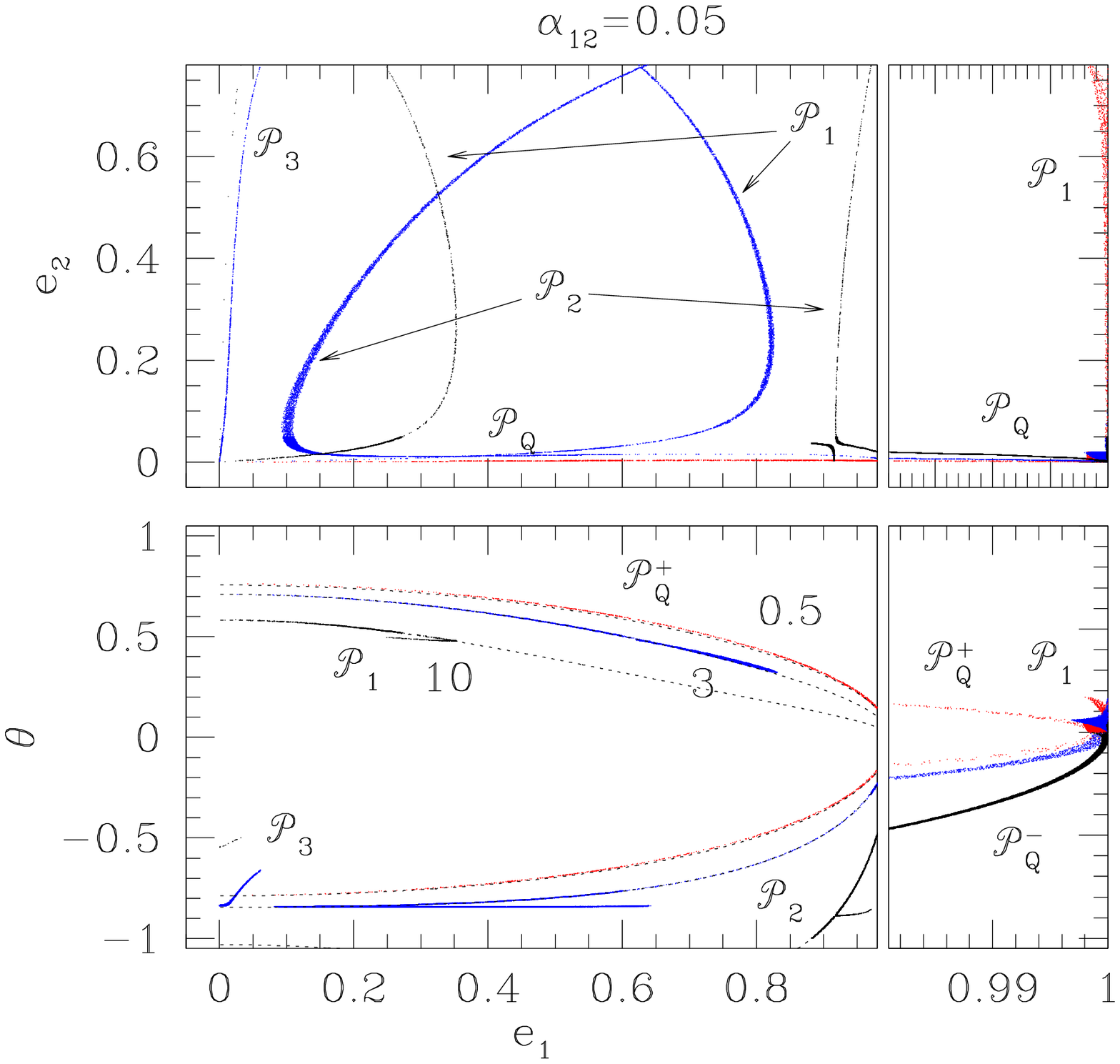}
                \caption{ This figure shows the stationary point families in the case $\omega_1=\omega_2=\pi/2$. 
The lower panel shows the relationship between $e_1$ and $\theta$ for different values
of the mass ratio. Families with  $\mu_{12}=0.5$ are shown in red, $\mu_{12}=3$ are shown in blue, and $\mu_{12}=10$ are
shown in black. The upper panel then shows the corresponding $e_1$--$e_2$ relationship. The right hand panels show
a zoom in on the $e_1 \sim 1$ region. The dotted curves indicate the corresponding quadrupolar Kozai-Lidov family at 
each $\mu_{12}$.
}
                \label{fig:evolperp}
        \end{figure}

Figure~\ref{fig:evolperp} shows the different solutions in case $\mathcal{P}$, as a function of mass. Once again, all solutions
are shown for $\alpha_{12}=0.05$. In red, we show the solutions for $\mu_{12}=0.5$ (lower mass ratios are qualitatively similar). Most obvious is a family of solutions that corresponds closely to the quadrupolar family $Q_2$ -- the Kozai-Lidov family. As in the quadrupolar case, this family becomes increasingly asymmetric with increasing $\mu_{12}$ and so we refer separately to $\mathcal{P}_{\rm Q}^+$ (prograde case) and $\mathcal{P}_{\rm Q}^-$ (retrograde case). As in the case of $\overline{\mathcal{A}}_{\rm Q}$, this family is found with
 small, but finite, $e_2$. 
 
 As the mass ratio increases,
 we also see the appearance of a second family, which we term $\mathcal{P}_1$. Examining the blue curves in Figure~\ref{fig:evolperp}, we see that the $\mathcal{P}_1$ family  appears to track $\mathcal{P}_{\rm Q}^+$ quite closely in terms of $e_1$--$\theta$, but deviates strongly in the upper panel, where it shows solutions 
 with a much larger $e_2$.  
We see also that the $\mathcal{P}_{\rm Q}^+$ family has a
maximum  $e_1$, which is also the point at which this
low $e_2$ family merges into the higher $e_2$ family $\mathcal{P}_1$. We see also  that the maximum value
decreases as $\mu_{12}$ increases.


The retrograde family, $\mathcal{P}_{\rm Q}^-$ also exhibits a higher $e_2$ counterpart which we term family $\mathcal{P}_2$, as can be seen in Figure~\ref{fig:evolperp}. This forms initially for small $e_1$ and moves to larger $e_1$ as $\mu_{12}$ increases. Once again we see that the quadrupolar analogue solution $\mathcal{P}_{\rm Q}^-$ merges smoothly with the higher eccentricity $\mathcal{P}_2$ family.

Finally, we also find a fourth family, a retrograde family we call $\mathcal{P}_3$. Both $\mathcal{P}_2$ and $\mathcal{P}_3$ emerge as the $e_1$ lower
limit of family $\mathcal{P}_{\rm Q}^-$ moves away from the circular orbits.

\subsection{Case $\overline{\mathcal{P}}$: $\omega_1=3\pi/2$, $\omega_2=\pi/2$}

\label{sec:CaseD}

\begin{figure}
                \centering
                \includegraphics[width=1.0\linewidth]{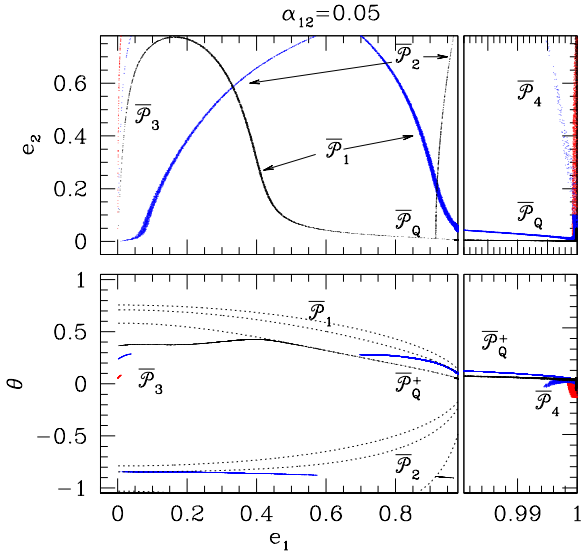}
                \caption{ This figure shows the stationary point families for the case $\omega_1=3\pi/2$ and 
$\omega_2=\pi/2$. The lower panel shows the relationship between $e_1$ and $\theta$ for different values
of the mass ratio $\mu_{12}$ -- $\mu_{12}=0.7$ is shown in red, $\mu_{12}=3$ is shown in blue, and $\mu_{12}=10$ is
shown in black. The upper panel then shows the corresponding $e_1$--$e_2$ relationship. The right hand panels show
a zoom in on the $e_1 \sim 1$ region. The dotted curves indicate the corresponding quadrupolar Kozai-Lidov family at
each $\mu_{12}$.
}
                \label{fig:antiperp}
        \end{figure}

Case~$\overline{\mathcal{P}}$ refers to the same equations~(\ref{eqn:ifix3}) and (\ref{eqn:ifix4}), but with the lower sign on the right hand side of each (positive and negative, respectfully). The stationary point families are shown in Figure~\ref{fig:antiperp}. As one might expect, there is a fair amount of symmetry between the solutions in Figure~\ref{fig:antiperp} and those in Figure~\ref{fig:evolperp}.

Perhaps the first point to note is the complementarity between these solutions and the corresponding apsidally aligned ones in Figure~\ref{fig:evolperp}.
Family $\famD_{\rm Q}$$^+$ appears to be the complement of $\famC_{\rm Q}$$^+$ -- occurring for those values of $e_1$ and $\theta$ where the 
solution is not found in case $\mathcal{P}$. Together, they appear to comprise the full analogue of the $Q_2^+$ solution.
There does not seem to be an equivalent $\overline{\mathcal{P}}_{\rm Q}$$^-$,
but that is not surprising given that $\mathcal{P}_{\rm Q}^-$  appears to cover the full range of eccentricities. In \S~\ref{sec:switch} we will show that the $\famD_{\rm Q}$$^-$ family does exist, but covers only a very limited range of parameters.

 We also find analogues of the high $e_2$ extensions of the quadrupolar analogue
families in $\overline{\mathcal{P}}_1$ and $\overline{\mathcal{P}}_2$ -- the equivalents of $\mathcal{P}_1$ and $\mathcal{P}_2$.
In addition, there exists a family $\overline{\mathcal{P}}_3$, an analogue to $\mathcal{P}_3$, but this time it occurs for prograde, rather than retrograde, configurations. For large enough masses ($\mu_{12}>10$), the families $\famD_{\rm Q}$$^+$, $\overline{\mathcal{P}}_1$ and $\overline{\mathcal{P}}_3$ form a continuous curve. 

Finally, the right hand panels of Figure~\ref{fig:antiperp} show a family, $\overline{\mathcal{P}}_4$ of almost radial orbits. Superficially,
these look like the mirror image of $\mathcal{P}_1$ at low masses. However, as $\mu_{12}$ increases, $\overline{\mathcal{P}}_4$ does not extend along
$\mathcal{P}_{\rm Q}^-$, but eventually retreats towards the radial limit again. Mass ratios of $\mu_{12} \sim 3$--5 
(for $\alpha=0.05$) mark the maximum extension of this family to smaller $e_1$ (which remains well above 0.9 at all times). This family is
also distinct from the $\overline{\mathcal{P}}_R$ family in the limit of $e_1 \rightarrow 1$, because it is found with
$e_1$ demonstrably less than unity (although still large). This family is characterised by large values in both the quadrupolar and octupolar terms in $\dot{\omega}_2$, which offset each other. The limited range of applicability in $\mu_{12}$ is a consequence of the $\epsilon<0.1$ cutoff -- this family extends over a much larger range of masses if we relax this criterion.

\subsection{Special Cases: Octupole limit}
\label{sec:OctoSpecial}

In the quadrupole limit, we also found that we could satisfy $\dot{e}_1=0$ and $\dot{\omega}_1=0$ in special
limiting cases, where $\omega_1$ was not restricted to the same values as in other stationary point families.
With the introduction of the octupole term, these families become even more restrictive, because the vanishing
of this term imposes conditions beyond those imposed by the quadrupole term. We will denote the special case
octupolar families with $\mathcal{O}$.

\subsubsection{Circular Limit: $\mathcal{O}_C$}

In the case of the circular limit $e_1=0$ (see \S~\ref{sec:Qc}), the octupole level introduces a term
$\dot{\omega}_1 \propto 1/e_1$, so this will only be zeroed if the coefficient of this term goes to zero simultaneously. This imposes a relationship between $\omega_1$, $\omega_2$ and $\theta$, as a function of $\mu_{12}$, $\alpha_{12}$ and
$e_2$. The condition $\dot{\omega}_2 = 0$ reduces to the quadrupolar limit in this case.

In the low mass ratio limit, this leads to $\theta=0$, $\omega_2=\pi/2$, and $\cos 2 \omega_1 = 1/5$ (i.e. $\omega_1=39.23^{\circ}$). 
 In the high mass limit, this requires $\theta=\pm 1/\sqrt{5}$ and one of $\omega_2=0$,  $\omega_1=0$ or $\omega_1=\pi/2$. 
These are the critical inclinations $I_{\rm crit}$ identified by \citep{JM66,Kras72}.

\subsubsection{Coplanar Limit: $\mathcal{O}_{\rm LL}$ and $\mathcal{O}_{||}$.}

\begin{figure}
                \centering
                \includegraphics[width=1.0\linewidth]{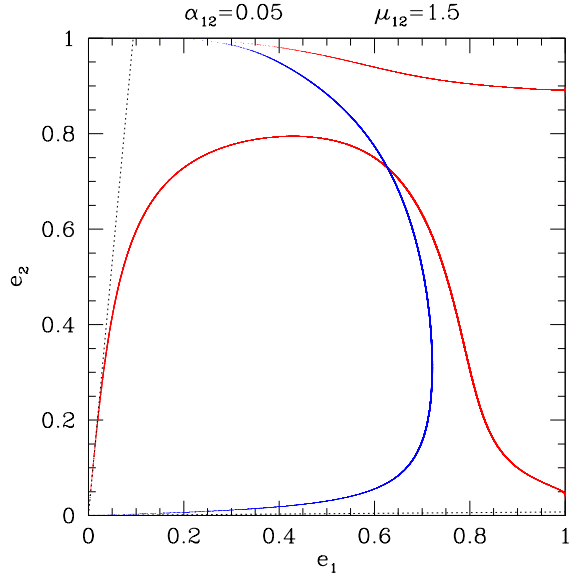}
                \caption{ The red curve shows the apsidally aligned, prograde coplanar fixed point solution, while the blue curve shows the apsidally anti-aligned case. The two dotted lines indicate the expected low eccentricity asymptotes for this case from the Laplace-Lagrange approximation. In this limit, equation~(\ref{eqn:LL}) reduces to a quadratic equation in $e_2/e_1$. In both apsidally aligned and anti-aligned cases, one of these roots is positive and these yield the solutions plotted here.
}
                \label{fig:Coplanar}
        \end{figure}

In the quadrupolar limit, we also found family $Q_{||}$ (see \S~\ref{sec:Qplanar}) in the limit of coplanar, retrograde orbits. So, let us now examine the case of $\theta=\pm 1$. In this case, we find that $\dot{e}_1=0$ and $\dot{e}_2=0$ are satisfied by the condition $\omega_1=\omega_2$. In the prograde case ($\theta=+1$), this relation can be maintained by $\dot{\omega}_1=\dot{\omega}_2$, because the equation derived from this condition is
\begin{eqnarray}
&& 2 \left( 1 - e_1^2 \right) - \left( 2 + 3 e_1^2 \right) \beta = 
 \frac{5}{8} \frac{\alpha_{12}}{1 - e_2^2} \cos \left( \omega_1 - \omega_2 \right) \times \nonumber \\
&& \left[ (1-e_1^2)(4+9 e_1^2) \frac{e_2}{e_1} - \beta (1+4 e_2^2) (4 + 3 e_1^2)
\frac{e_1}{e_2} \right], \label{eqn:LL}
\end{eqnarray}
which depends only on the angle $\omega_1 - \omega_2$. This is the extension of the traditional Laplace-Lagrange treatment (e.g. \cite{MD99}) to the hierarchical case \citep[e.g.][]{LP03,MM04}. As in the traditional case, this yields two solutions
corresponding to $\omega_1-\omega_2=0$ or $\pi$, although the nonlinearity of the system means that these no longer form a basis set for describing more general behaviour. Figure~\ref{fig:Coplanar} shows an example solution. Both curves asymptote to a fixed value of $e_2/e_1$ at small eccentricities, which matches the expectations from the traditional expansion. We will refer to this family as '$\mathcal{O}_{\rm LL}$', since it represents the extension of the Laplace-Lagrange family. Note that this analogue of the classical family  appears first at octupole order, since it depends on the difference between $\omega_1$ and $\omega_2$.

In the retrograde case, the equation for $\dot{\omega}_1 - \dot{\omega}_2$ depends on $\omega_1$, and so it is only a stationary point if $\dot{\omega}_1=0$ as well. This implies $\dot{\omega}_2=0$ also, i.e. the same fixed point condition as in the other cases. From this 
 we derive a relationship between $\omega_1$ and $e_1$ that is an extended version of equation~(\ref{eqn:quadcop}) which yields a qualitatively similar solution -- we find sensible solutions only in the retrograde case and for sufficiently large $\mu_{12}$. In the online appendix,Figure~B10 shows an example of the solution
for the case $\mu_{12}=10$ and $\alpha_{12}=0.05$. The shape of the solution closely tracks the quadrupolar version, but there is a change of apsidal alignment along the curve. Comparison with the other solutions shows that this family forms the same kind of link between $\overline{\mathcal{A}}_{\rm Q}$ and $\mathcal{P}_{\rm Q}^-$ as the corresponding quadrupolar solution does.
 We will call this family $\mathcal{O}_{||}$.


\subsubsection{Radial limit: $\mathcal{O}_{\rm R}$}
\label{sec:ORlimit}

Finally, we have the radial limit $e_1 \rightarrow 1$. In this limit, $\dot{e}_1=0$ automatically, and $\dot{\omega}_1=0$
imposes additional constraints, given by the condition
\begin{eqnarray}
 && \theta^2 \left(1 - \cos 2 \omega_1\right) = - \frac{105}{12} \alpha_{12} \theta \frac{e_2}{1-e_2^2} \times \nonumber \\
&& \left[ \sin \omega_1 \sin \omega_2 
. \left( 1 + \frac{5}{2} (1 - \cos 2 \omega_1) (3 \theta^2 - 1) \right)
+ \right. \nonumber \\
&& \left. 5 \theta (1 - \cos 2 \omega_1) \cos \omega_1 \cos \omega_2 \right]. \label{eqn:OReqn}
\end{eqnarray}
 This is automatically satisfied if $\theta=0$ or $\omega_1=0$ but more general combinations of $\omega_1$, $\omega_2$ and $\theta$ also satisfy this criterion. 
 The addition of the constraint that $\dot{\omega}_2=0$ as well does not restrict the solutions because $\dot{\omega}_1=\dot{\omega}_2$ in this limit. 
This leads to a rather broad family of possible solutions, which we term $\mathcal{O}_{\rm R}$.

\begin{figure}
                \centering
                \includegraphics[width=1.0\linewidth]{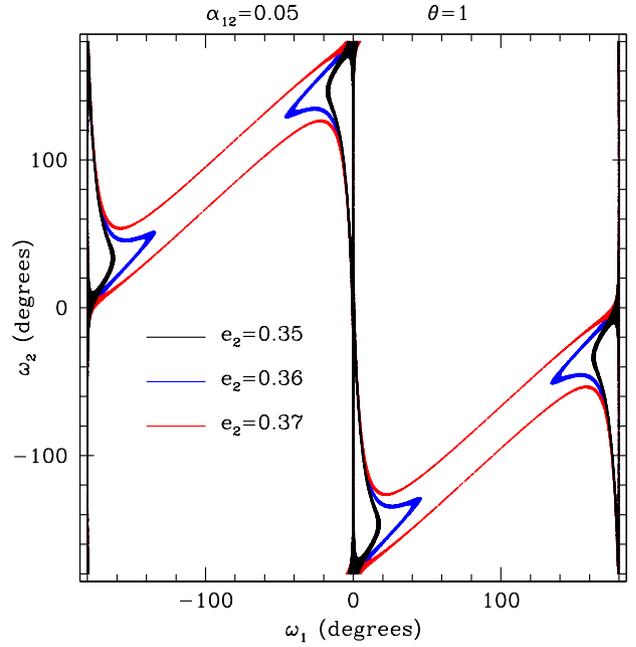}
                \caption{Each curve represents the solution to equation~(\ref{eqn:OReqn}) for the case $\theta=1$.
The vertical lines at $\omega_1=0$, $\pi$ and $2 \pi$ represent the fact that the equation is satisfied for all $\omega_2$ at these values. For other values of $\omega_1$, the equation implies a relationship between $\omega_2$ and $e_2$ for fixed $\alpha_{12}$. We see that this more general solution is narrowly confined to near the special values of $\omega_1$ as long as $e_2$ is below a threshold value. However, as $e_2$ increases above this value (which has the value $e_2=0.361$ for this case) we see that the topology of the solution changes dramatically.
}
                \label{fig:Spec3}
        \end{figure}

An important potential application of this family is in the case of coplanar orbits ($\theta \rightarrow \pm 1$). In that instance, the more general class of solutions are clustered near $\omega_1=0$ or $\pi$, unless $e_2$ is above
some threshold value. We can estimate the critical $e_2$ by setting $\omega_1=\pi/2$ and deriving the resulting $\omega_2$ from
\begin{equation}
\sin \omega_2 = \mp \frac{24}{1155} \frac{(1-e_2^2)}{e_2 \alpha_{12}}. \label{eqn:ORcrit}
\end{equation}
The requirement that $\left| \sin \omega_2 \right|<1$ imposes a condition on $e_2$. For $\alpha_{12}=0.05$, this
is $e_2>0.361$. Figure~\ref{fig:Spec3} shows the nature of the solution near this critical value. We see that the
shift is quite dramatic, over only $\Delta e_2=0.02$. Note that the form of equation~(\ref{eqn:ORcrit}) is the
same as the expansion parameter $\epsilon=\alpha_{12} e_2/(1-e_2^2)$, so that the critical value can also be
expressed as a critical $\epsilon_{crit}=0.021$. This becomes relevant in the case of coplanar orbital flips, as discussed in \S~\ref{sec:CopFlip}.

\section{Effect of Relativity}
\label{sec:Relative}

For planets in short period orbits, the effect of relativistic precession must be included. The addition of a component of $\dot{\omega}_1$ from relativity 
can shift the location of these stationary points. Indeed, relativistic effects can give rise to qualitatively new features in the case of massive outer perturbers or comparable mass inner binaries 
\citep[e.g.,][]{Naoz+13,NLZ17,Naoz+20,Will14,Will17,Liu+15,Liu+19,Lim+20}, but, with our focus on the planetary case, we will limit our attention to the post-Newtonian correction to the inner orbit precession.

Relativistic precession is usually discussed in the context of the suppression or excitation of orbital eccentricity, but it can shift both the inclination and eccentricity of the stationary points \cite[e.g.][]{MG11}. To illustrate this, let us
consider the addition of a relativistic contribution (equation~57 of \cite{Naoz16}) to the
right hand side of equation~(\ref{eqn:quadwdot}). If we set $\dot{\omega}_1=0$ in the case
of $\omega_1=\pi/2$ (the $Q_2$ solution), we now derive a modified condition on the 
inclination of the fixed point, namely 
\begin{equation}
\theta = \theta_0 \left[ -1 \pm \left( 1 + \frac{60 ( 1 - e_1^2) - \gamma}{\mu_{12}^2 \alpha_{12} (1 - e_1^2) (1 + 4 e_1^2)^2}\right)^{1/2} \right], \label{eqn:RelQ2}
\end{equation}
where $\theta_0 = 0.1 \mu_{12} \alpha_{12}^{1/2} (1 - e_1^2)^{1/2} ( 1 + 4 e_1^2)$ and
\begin{equation}
\gamma = 40 \frac{M_c}{M_2} \frac{R_S a_2^3}{a_1^4} \frac{ ( 1 - e_2^2)^{3/2}}{(1-e_1^2)^{1/2}}, 
\end{equation}
and $a_1$, $a_2$ are the semi-major axes of the inner and outer planets, and $R_S = 2 G M_c/c^2$ is the Schwarzschild radius of the  central object. This is essentially a `squashed' version of the original $Q_2$ family, since e$_1$ is now bounded from above at a smaller value than unity. 
The equivalent correction for the $Q_1$ family shifts the mass threshold at which it appears. Note also that the relativistic contribution has broken the scale invariance of the problem, since lengths are now scaled relative to $R_S$.

\begin{figure}
                \centering
                \includegraphics[width=1.0\linewidth]{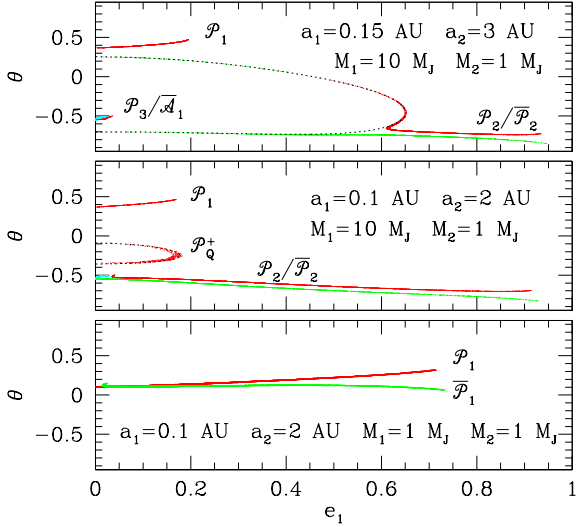}
                \caption{The upper panel shows the fixed point families for the case of $\alpha_{12}=0.05$, $\mu_{12}=10$, where the scales are set by $a_1=0.15$AU and $M_1=10 M_J$. The dotted line is the `squashed' $Q_2$ family given by equation~(\ref{eqn:RelQ2}). The middle panel shows the effect of shifting everything inwards so $a_1=0.1$AU. The $Q_2$ family is now entirely retrograde and the $\mathcal{P}_2$ and $\overline{\mathcal{P}}_2$ families have now detached from $\mathcal{P}_{\rm Q}^+$. The bottom panel shows the result of keeping a$_1$ fixed but reducing $M_1$. We see that the $\mathcal{P}_{\rm Q}^+$ family completely disappears at this point, leaving only $\mathcal{P}_1$ and $\overline{\mathcal{P}}_1$. The $\overline{\mathcal{P}}_1$ family was not present in the upper two panels and only appears when $M_1<6 M_J$, in this case.
}
                \label{fig:3Pan}
        \end{figure}

Figure~\ref{fig:3Pan} shows the effect of relativity on the fixed point families of the full octupole problem, as we move the system closer to the star, while keeping $\alpha_{12}=0.05$~fixed. The upper two panels show the case of $\mu_{12}=10$ (realised in this case by
$M_1=10 M_J$ and $M_2=1 M_J$). Far from the star, the fixed point families should look as they do in Figure~\ref{fig:evolperp} and \ref{fig:antiperp}. However, the upper panel shows that, if $a_1=0.15$AU, then the effects of relativity are significant. The quadrupolar family is squashed and distorted, and the positions of additional families (such as $\mathcal{P}_1$ or $\mathcal{P}_2$) are shifted as well. The middle panel shows the effect of moving the system in even further ($a_1=0.1$AU). We see now that the $\mathcal{P}_2$ and $\overline{\mathcal{P}}_2$ families no longer intersect the quadrupolar family (which is now squashed down to e$_1$<0.2). 

The bottom panel of Figure~\ref{fig:3Pan} shows what happens if we decrease the mass ratio to $\mu_{12}=1$ (M$_1=1 M_J$) at this location. Now the quadrupolar family has completely disappeared \citep[see also][]{FKR00,Naoz+13,Naoz+20}, but the $\mathcal{P}_1$ and $\overline{\mathcal{P}}_1$ families remain. This illustrates that the inner planet mass is important too.

   \begin{figure}
                \centering
                \includegraphics[width=1.0\linewidth]{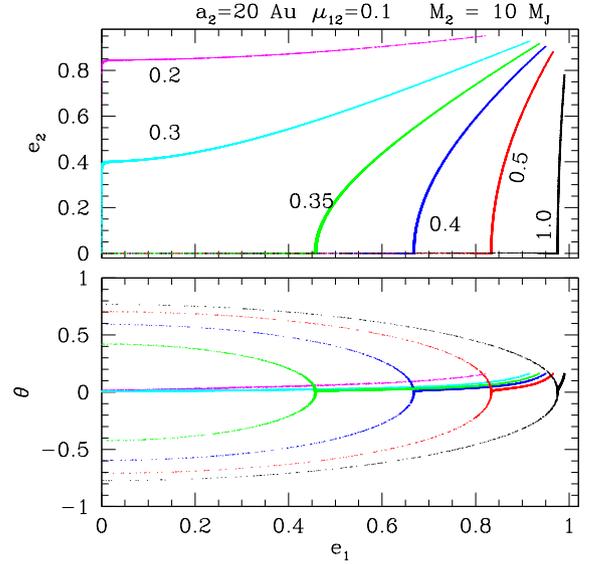}
                \caption{These curves illustrate the effect of relativistic precession on the fixed point families for the case of $\omega_1=\omega_2=\pi/2$. The outer planet semi-major axis is kept fixed at 20~AU, but we show here the fixed point families for a variety of inner semi-major axes -- 1 AU (black), 0.5 (red), 0.4 (blue), 0.35 (green), 0.3 (cyan) and 0.2~AU (magenta). The e$_1$--e$_2$ relation in the upper panel demonstrates that there are two components -- an essentially quadrupolar version (e$_2 \sim 0$) that looks like a `squashed' version of the $Q_2$ family, and an extension of the
$\mathcal{P}_1$ family that extends to large e$_1$ and e$_2$.
}
                \label{fig:Rel3}
        \end{figure}

One of the reasons to be interested in the effect of relativity is in the case of planet migration driven by secular interactions and tides. In this case, we should keep the outer planet location fixed ($a_2$ fixed, rather than $\alpha_{12}$) and move $a_1$ inwards. Thus, 
as $a_1$ decreases, for fixed $a_2$, the $\gamma$ contribution increases more rapidly. When the numerator factor in equation~(\ref{eqn:RelQ2}) equals zero, $\theta=0$ and this is the maximum $e_1$ for which the quadrupolar solution remains, namely
\begin{equation}
e_1 = e_{max} =  \left[ 1 - \left( \frac{2 M_c}{3 M_2} \right)^{2/3} \frac{R_S^{2/3} a_2^2}{a_1^{8/3}} \right]^{1/2}.
\end{equation}
Eventually, $a_1$ is small enough that $e_{max}=0$, which yields the criterion for the Kozai-Lidov family to survive:
\begin{equation}
a_1 > 0.32 AU \left( \frac{a_2}{20 AU} \right)^{3/4} \left( \frac{M_2}{ 0.01 M_{\odot}} \right)^{-1/4}
\left( \frac{M_c}{M_{\odot}} \right)^{1/2}. \label{eqn:KLsurvive}
\end{equation}

Figure~\ref{fig:Rel3} shows the evolution of the full fixed point family (to octupolar order) in an example where
we hold the outer perturber fixed at 20~AU, but move the inner planet closer to the star. We show here the stationary
point families in the case
of
$\omega_1=\omega_2=\pi/2$. We see the evolution of this `squashed' Kozai-Lidov family to ever smaller $e_{max}$
as $a_1$ decreases. However, we see that the $\mathcal{P}_1$ family remains and plays an ever larger role as $e_{max}$
decreases. This family starts at $e_{max}$ and extends up to almost radial orbits. The family is also strongly
polar. Even when the $Q_2$ family disappears, the $\mathcal{P}_1$ family continues to exist for non-zero e$_2$. This eventually
disappears too at $\sim 0.15$AU (for this example). The equivalent anti-aligned case ($\overline{\mathcal{P}}_1$) shows a similar form but
extends to slightly retrograde orbits instead of slightly prograde ones.

The $\omega_1=0$ cases remain qualitatively similar with the introduction of relativity. Of more interest are
the special cases, because the allowed values of $\omega_1$ are shifted by the relativistic precession,
and because tidal circularisation naturally takes us to the $e_1 \rightarrow 0$ limit. 
The one that survives the furthest in is the polar $\mathcal{O}_{\rm C}$ limit, wherein
\begin{equation}
\cos 2 \omega_1 = \frac{1}{5} \left[ 1 - \frac{8 R_S a_2^3}{a_1^4} \frac{M_c}{M_2} \left(1-e_2^2\right)^{3/2}.
\right]
\end{equation}
For this to provide a physically reasonable answer, $\cos 2 \omega_1 > -1$, which implies
\begin{equation}
a_1 > 0.278 AU \left( \frac{a_2}{20 AU} \right)^{3/4} \left( \frac{M_2}{0.01 M_{c}}\right)^{-1/4}
\left( 1 - e_2^2 \right)^{3/8}. \label{eqn:rellimit}
\end{equation}
Comparison with equation~(\ref{eqn:KLsurvive}) shows that the $\mathcal{O}_C$ saddle point survives longer than the Kozai-Lidov quadrupolar fixed point family as a planet is dragged down by tides.

Thus, the effects of relativity start to have a marked effect on the stationary point families for $a_1 < 1$AU. In
the case where we keep $a_2$ fixed -- so that $\alpha_{12}$ decreases as $a_1$ does, 
 the bulk of the stationary points are wiped out by $a_1 \sim 0.3$AU. For more compact systems, wherein we keep $\alpha_{12}$ fixed as we move the planet pair inwards, some stationary points can survive interior to 0.1~AU. A generic tendency is for the surviving stationary point families to lie close to polar. 

\section{Stability of the Stationary Point families}
\label{sec:Stability}

In the previous sections, we reviewed the stationary point solutions of the octupolar-level expansion of the hierarchical three-body problem. This identifies equilibria, but does not specify the stability of said equilibria. Here we shall review each family of stationary points in order to determine their stability -- whether they are fixed points or saddle points -- and their role in fixing the orbital structure. Furthermore, we will do this as a function of $\mu_{12}$. In the previous section we arranged our solutions in terms of their apsidal alignment, but it is also instructive to see how the various classes fit together for particular mass ratios.
We will once again fix $\alpha_{12}=0.05$, so that we can vary $\mu_{12}$ alone. The results are summarised in Table~\ref{StableTable}. 

To determine stability, we will make use of the fact that the short-term dynamics are still driven primarily
by the quadrupolar term \citep{NF13,LN14,Antognini15,Naoz16}, with longer-term drifts imposed by octupolar contributions. So, we first construct contours of constant energy and angular momentum, to identify whether the underlying short-term dynamics are consistent with a fixed point or a saddle point. We then follow this with direct integrations of the full octupolar equations to verify whether the octupolar terms change the long-term dynamics. 

In the octupolar case, we also encounter families where $\omega_1$ librates about the equilibrium while $\omega_2$ displays a saddle point behaviour. In this case $\omega_2$ circulates, but can show a brief reversal that qualifies as a solution to our equilibrium conditions of $\dot{e}_1=\dot{e}_2=\dot{\omega}_1=\dot{\omega}_2=0$. In this case we will refer to an `inner fixed point'.

\subsection{Low mass ratios: $\mu_{12}=0.1$}

Empirically, this is the most common kind of system observed in exoplanet systems. 
It is also the least complicated case, because the small inner mass induces a limited precession of the outer
mass and so the configuration of the stationary point families hews pretty close to the quadrupolar case.
Figure~B5 of the online appendix summarises the stationary point families present.


\subsubsection{Kozai-Lidov Analogue: $\mathcal{P}_{\rm Q}$}

As one might expect, this limit is dominated by the $\mathcal{P}_{\rm Q}$ and $\overline{\mathcal{P}}_{\rm Q}$ families, the generalisation of the Kozai-Lidov $Q_2$ family.
However, despite the low $\mu_{12}$, both $\omega_1$ and $\omega_2$ librate  so these represent a full generalisation of the Q$_2$ families to the octupolar case. Figure~B6 of the appendix shows an example of such a trajectory.

\subsubsection{Saddle Points: $\mathcal{A}_1$}

The $\mathcal{A}_1$ family is also present in the low $\mu_{12}$ limit, where it occurs for almost polar orbits.
This is a saddle point. A trajectory that begins near this point exhibits large-scale variations in $e_1$, and is
located at the extreme of librations about the $\mathcal{P}_{\rm Q}$ fixed point. This stationary point performs the
same role as the $Q_C$ fixed point in the quadrupole description (see \S~\ref{sec:Qc} and Figure~\ref{fig:Example}), but
is more localised because it includes the criterion that $\dot{\omega}_2=0$, which restricts the family to almost 
polar configurations. It is also worth noting that many of these trajectories yield intermittent orbital flips,
as $i_{\rm tot}$ can fluctuate about the polar value.


\subsubsection{Radial Families}
\begin{figure}
                \centering
                \includegraphics[width=1.0\linewidth]{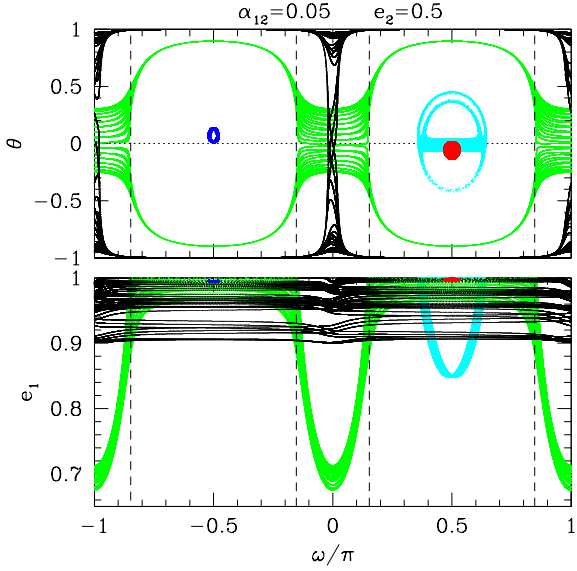}
                \caption{ The lower panel shows the evolution of $e_1$ with $\omega_1$ for five different orbital integrations. All integrations start with $e_1=0.995$, $e_2=0.5$, $\theta=0$, $\omega_2=\pi/2$. The integrations shown in red start with $\omega_1=\pi/2$ and the blue integrations with $\omega_1=-\pi/2$. These correspond to the radial families  $\mathcal{P}_{\rm R}$ and $\overline{\mathcal{P}}_{\rm R}$,  and demonstrate the stability of the equilibrium in this limit.
 The integrations shown in black begin with 
$\omega_1=\pi$. This shows the chaotic switching between prograde and retrograde orbits discussed by \citet{LN14,LNH14L} and 
 \citet{Naoz16} and references therein. Note also that this trajectory spends a lot of time in a coplanar configuration -- either in prograde or retrograde directions.
 The green integrations start with $\omega_1=26^{\circ}$, which is close to the special case of $\mathcal{O}_{\rm R}$ discussed in the text. We see that this is a saddle point -- once the value of $\omega_1$ approaches the critical value (shown by the vertical dashed lines), it transitions from libration to circulation. This family is therefore the saddle point that separates the libration and circulation regimes of $\omega_1$
 The cyan integrations start with $\omega_1=66^{\circ}$ -- an intermediate value. In this case, we see $\omega_1$ still librates, but experiences flips from prograde to retrograde, illustrating that orbital flips do not require passage through a saddle point in $\omega_1$. 
}
                \label{fig:CR_low}
        \end{figure}

One feature to note, in addition to the analogues of the quadrupolar families, is the presence of the quasi-radial stationary point
families $\mathcal{P}_{\rm R}$, $\overline{\mathcal{P}}_{\rm R}$ and  $\mathcal{O}_{\rm R}$, as shown in Figure~\ref{fig:CR_low}. To demonstrate their role, we choose a starting point close to $\theta \sim 0$, and integrate the equations for different initial apsidal misalignments (varying $\omega_1$ keeping $\omega_2$ fixed).
 Although the $\mathcal{P}_{\rm Q}$ curves pass through the same point in $e_1$--$\theta$ space, these $\mathcal{O}_{\rm R}$ equilibria exist for large $e_2$ as well. The examples shown in Figure~\ref{fig:CR_low} are integrated using an initial $e_2=0.5$. 

These integrations establish that $\mathcal{P}_{\rm R}$ (red points) and $\overline{\mathcal{P}}_{\rm R}$ (blue points) are examples of the inner fixed point variety discussed above --
the angle $\omega_2$ does not librate over a finite range but circulates, while $\omega_1$ librates. 
  The 
 circulation of $\omega_2$ introduces a small variation in the parameters that can generate orbital flips if the librations are large enough to approach $\omega_1 \sim 0$ or $\pi$. However, this requires large amplitude librations  -- we have not found cases where the small amplitude librations are destabilised, because  the variation in $e_2$ is not large for low $\mu_{12}$. If we start with enough apsidal misalignment, we do indeed see  orbital flips, as shown by the cyan points. 
 These integrations also establish the relationship between the different radial families. The $\mathcal{P}_{\rm R}$ and $\overline{\mathcal{P}}_{\rm R}$ families are inner fixed points and the $\mathcal{O}_{\rm R}$ family is a saddle point that separates the regimes of libration and circulation of $\omega_1$, as shown by the green and black points.

\subsection{Comparable mass ratios: $\mu_{12}=1.5$}

As the mass ratio becomes comparable, the effects of the inner planet on the outer become stronger, and
start to introduce features not found in the limit of an inner test particle.
Figure~\ref{fig:medmu} shows the different stationary point families for the case $\mu_{12}=1.5$.
The four different cases for $\omega_1$ and $\omega_2$ are shown in blue (Case $\mathcal{A}$), cyan (Case $\overline{\mathcal{A}}$),
red (Case $\mathcal{P}$) and black (Case $\overline{\mathcal{P}}$).

\subsubsection{Kozai-Lidov Generalisations: $\famC_{\rm Q}$ and $\famC_1$}
We see that the most prominent feature is still the generalised 
version of the Kozai-Lidov family ($\mathcal{P}_{\rm Q}^+$ and $\mathcal{P}_{\rm Q}^-$). 
We see also the emergence of the $\mathcal{P}_1$
family, the eccentric version of the K-L family.


       \begin{figure}
                \centering
                \includegraphics[width=1.0\linewidth]{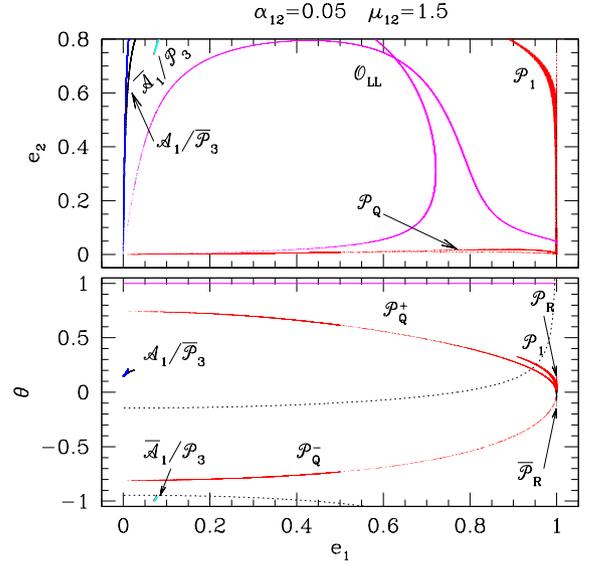}
                \caption{ The lower panel shows the $e_1$--$\theta$ relations for each of the stationary point
families observable in this case. The upper panel shows the corresponding $e_1$--$e_2$ relationships. Only points
that satisfy $\epsilon<0.1$ are shown. Case $\mathcal{A}$ ($\omega_1=\omega_2=0$) is shown in blue. Case $\overline{\mathcal{A}}$ ($\omega_1=\pi$,
$\omega_2=0$) is shown as cyan. Case $\mathcal{P}$ ($\omega_1=\omega_2=\pi/2$) is shown in red, and Case $\overline{\mathcal{P}}$ ($\omega_1=3\pi/2$,
$\omega_2=\pi/2$) is shown in black. The $\overline{\mathcal{A}}_1$ and $\famC_3$ families lie exactly on top of one another, so that the cyan points in this figure cover a similar feature in red. The $\famO_{\rm LL}$ extension of the Laplace-Lagrange stationary points are
shown in magenta. The dotted lines represent two cases of fixed total angular momentum, $G_0^2=0.023$ (upper curve) and $G_0^2=-0.16$ (lower curve). In both cases, $e_2=0.73$.
  }
                \label{fig:medmu}
        \end{figure}

       \begin{figure}
                \centering
                \includegraphics[width=1.0\linewidth]{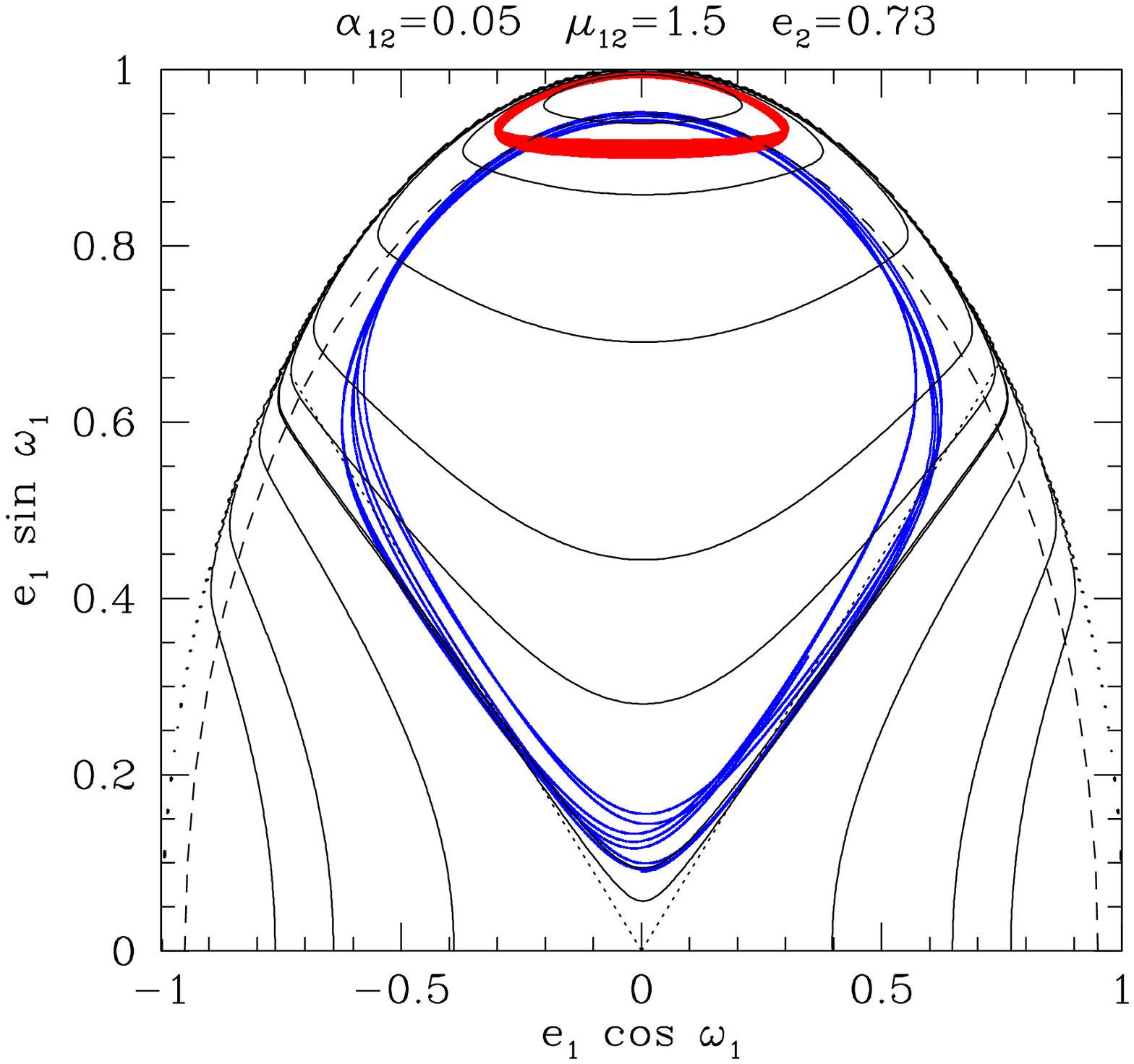}
                \caption{The solid curves are loci of constant energy, subject to a fixed total angular momentum,
given by the upper dotted line in Figure~\ref{fig:medmu}. The dashed line represents $e_1=0.95$, which is the expected
value of the stationary point $\famC_1$ identified for these parameters. We see that this passes through the centre of
libration. The red points indicate a direct integration, using $e_1=0.9$ and $\omega_1=\omega_2=\pi/2$ as initial
conditions. The dotted lines indicate the value of $\omega_1$ derived from the special solution $\famO_{\rm C}$.
The blue points show a trajectory that illustrates this family, which starts with the same initial conditions as
the red trajectory, except that $e_1=0.09$ initially.
}
\label{fig:MapC1}
\end{figure}

Figure~\ref{fig:MapC1} shows the contours of constant energy for the case where the angular momentum is given by the 
upper dotted line in Figure~\ref{fig:medmu}. We also need to specify a value of $e_2$, because this contour crosses both the $\famC_{\rm Q}^+$ and $\famC_1$ families, which have very different values of $e_2$ at their intersections. We know that the $\famC_{\rm Q}^+$ family is a stable point from the quadrupolar analysis, so we choose $e_2=0.52$ to isolate the $\famC_1$ stationary point.
 From the contours in Figure~\ref{fig:MapC1} we identify $\famC_1$ as a stable equilibrium -- a fixed point.
We confirm this by a direct integration of the orbital equations, shown in red. The finite width of the libration trajectory is a consequence of the small variation of $e_2$ (which is assumed to be constant in the calculation of the contours). Therefore, $\famC_1$ represents a high eccentricity offshoot of the $\famC_{\rm Q}$ family.

\subsubsection{The Prograde Saddle Points: $\famA_1$ and $\famD_3$}

The families $\famA_1$ and $\famD_3$ exist at low $e_1$
(for all $\mu_{12}$ and $e_2$). Direct integrations from the starting points of $\famA_1$ and $\famD_3$ families  indicate that these two are saddle points, representing the
minima of large amplitude librations or circulations. These saddle points are also associated with the special case
solutions $\famO_{\rm C}$. 
The dotted lines in Figure~\ref{fig:MapC1} indicate the angle appropriate for the special case $\famO_{\rm C}$ -- it is along
this angle that large amplitude librations approach the saddle point at the origin. The geometry of the curves in this
figure also illustrate the nature of the $\famA_1$ and $\famD_3$ stationary points -- the curves that turn away from the origin indicate that these are  saddle points. Inspection of Figure~\ref{fig:MapC1} might suggest that saddle points exist for all four possible apsidal alignments, but the requirement $\dot{\omega}_2=0$ restricts this to just the $\famA$ and $\famD$ cases. 

\subsubsection{The Retrograde Families $\famB_1$ and $\famC_3$}
\label{sec:C3}

\begin{figure}
                \centering
                \includegraphics[width=1.0\linewidth]{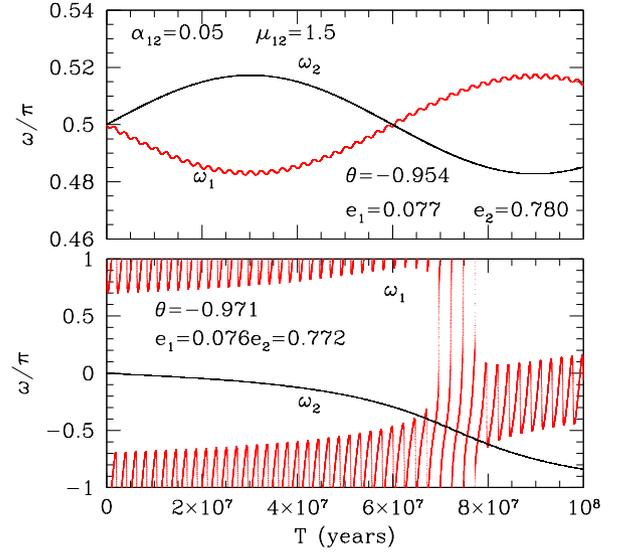}
                \caption{The upper panel shows the evolution for an example of starting conditions that
belong to the $\famC_3$ stationary point family. The starting values are $\omega_1=\omega_2=\pi/2$. We see that
this is a global equilibrium in the sense that both $\omega_1$ and $\omega_2$ librate. The shorter timescale libration is driven by the quadrupole potential, while the longer timescale variations are driven by the octupole.
 In the lower panel,
we show an example of the $\famB_1$ family. The eccentricities and mutual inclination are almost the same as the
upper panel, but the initial starting values are $\omega_1=\pi$ and $\omega_2=0$. In this case, $\omega_1$
librates with a drifting centre, and $\omega_2$ circulates (albeit slowly). The combination $\omega_1+\omega_2$ librates about
$\pi$, with brief periods of circulation when $\omega_2 \sim \pm \pi/2$.
}
\label{fig:C3evol}
\end{figure}

At the lower left in Figure~\ref{fig:medmu}, the cyan feature represents the appearance of the $\famB_1$ and $\famC_3$ retrograde orbital families.
The $\famC_1$ family appears to be a high eccentricity offshoot of the traditional Kozai-Lidov quadrupolar family, but the
 families $\famB_1$ and $\famC_3$ appear to be  qualitatively distinct. Figure~\ref{fig:C3evol} shows two
examples of direct integration - one chosen from the $\famC_3$ branch and one from the $\famB_1$ branch. The $\famC_3$ family is shown to be a stable fixed point family, with both $\omega_1$ and $\omega_2$ librating about the equilibrium values. The $\famB_1$ family, on the other hand, shows libration of $\omega_1$ on short timescales,
but an overall circulation of the libration centre on longer timescales, combined with circulation of $\omega_2$. We find that the combination $\omega_1+\omega_2$ librates about a value of $\pi$ in this case, which means that the two planets precess at roughly the same rate and maintain a maximal separation of their perihelia.
The value of $e_1$ also undergoes a large excursion with a minimum at the stationary point, so the $\famB_1$ family is most accurately characterised as a saddle point.
These two families are likely to be related to the known families of stable retrograde orbits in the problem of equal masses (e.g. \cite{Hen76}).

\subsubsection{The Laplace--Lagrange Analogues:}

For completeness, we note also the presence of the two $\famO_{\rm LL}$ families in this case, but do not discuss them
further as their behaviour and stability is well documented \citep{LP03,MM04} -- at least until they approach the orbit crossing limit.

\subsection{Large mass ratios: $\mu_{12}=10$}

\begin{figure}
                \centering
                \includegraphics[width=1.0\linewidth]{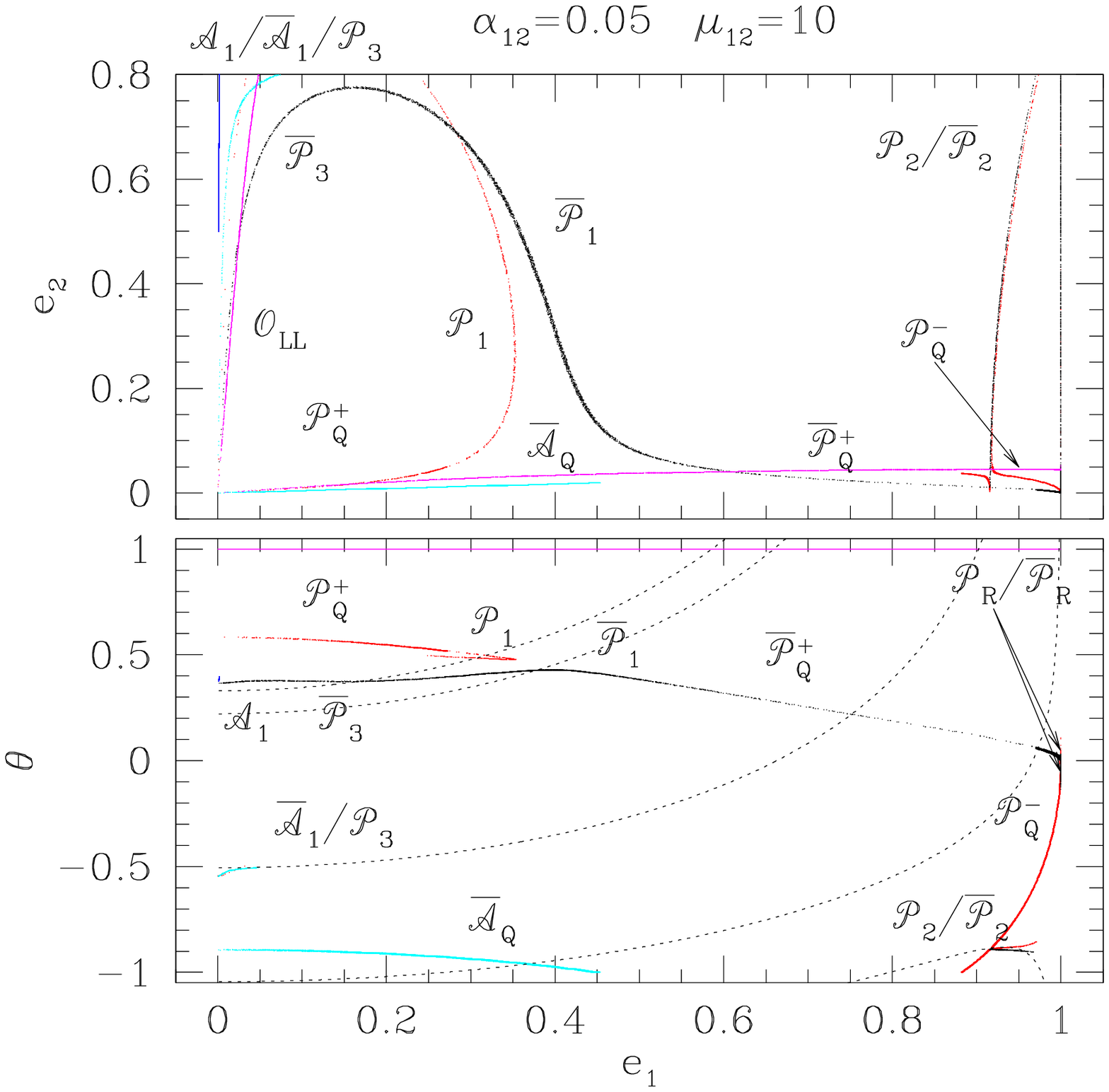}
                \caption{The lower panel shows the $e_1$--$\theta$ relations for the stationary point families in the
case of $\alpha_{12}=0.05$ and $\mu_{12}=10$. The upper panel shows the corresponding relations for $e_1$--$e_2$. The
colours indicate the apsidal geometries -- blue represents $\omega_1=\omega_2=0$, cyan represents $\omega_1=\pi$ and
$\omega_2=0$, red represents $\omega_1=\omega_2=\pi/2$ and black represents $\omega_1=3\pi/2$, $\omega_2=\pi/2$. 
The magenta curves represent the $\famO_{\rm LL}$ family -- the extension of the Laplace-Lagrange solutions. The gap between $\famB_{\rm Q}$ and $\famC_{\rm Q}^-$ is well matched by the $\famO_{||}$ solution shown in Figure~B10 in the online appendix. The dashed curves represent contours of constant angular momentum, and will be described in the following figures.
}
\label{fig:himufig}
\end{figure}

The landscape of stationary points gets more complicated as the mass ratio increases -- as is shown in
Figure~\ref{fig:himufig}. This is not surprising, as a more massive inner planet is more capable of
affecting the orbital dynamics of the outer planet.

\subsubsection{The Kozai-Lidov generalisations: $\famC_{\rm Q}$, $\famC_1$ and $\famC_2$: $\famD_{\rm Q}$, $\famD_1$ and $\famD_2$}

The asymmetry of the Kozai-Lidov $\famC_{\rm Q}$ family between prograde and retrograde is clearly evident in Figure~\ref{fig:himufig}. More interesting is the fact
that the prograde branch is split into an apsidally aligned $\famC_{\rm Q}^+$ ($e_1<0.4$) and apsidally anti-aligned $\famD_{\rm Q}^+$ ($e_1>0.4$) branch. Both also merge smoothly into their higher $e_2$ analogues $\famC_1$ and $\famD_1$.
 The $\famD_1$ family now also merges smoothly
with the $\famD_3$ family.
The retrograde family now shows a high eccentricity offshoot as well, with $\famC_2$ and $\famD_2$ now present at large $e_1$.


Figure~B7 of the online appendix shows the energy contours along the two uppermost of the dotted lines in Figure~\ref{fig:himufig}, representing two choices for the total angular momentum of the system and illustrates that the $\famC_1$ and
$\famD_1$ families still represent stable librations, as they did for lower mass ratios.



\begin{figure}
                \centering
                \includegraphics[width=1.0\linewidth]{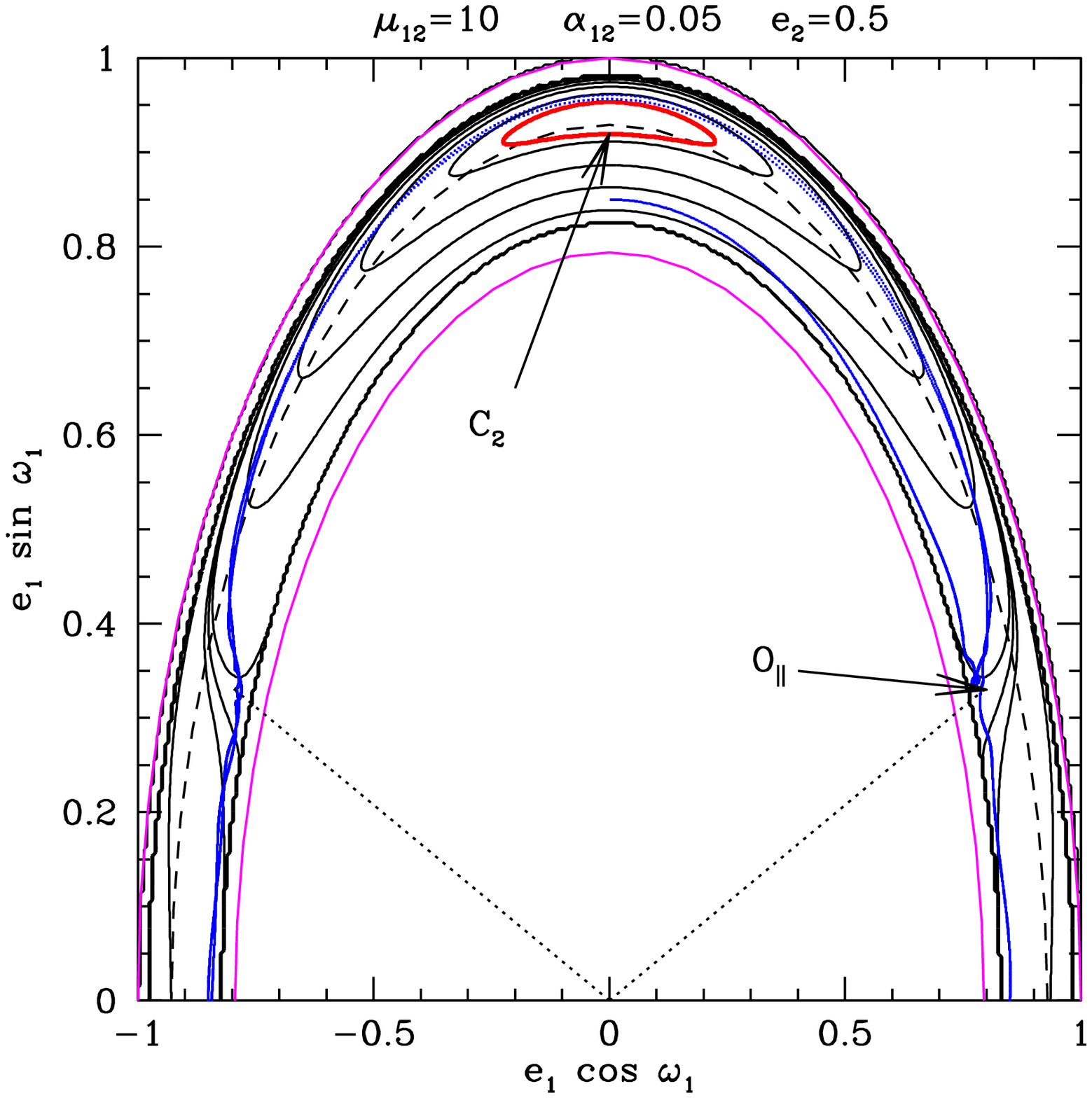}
                \caption{The contours represent constant energy at a fixed angular momentum given by the
dotted curve in the lower right hand corner of the lower panel in Figure~\ref{fig:himufig}, specifically for $G_0^2=-0.583$ and $e_2=0.5$. This represents
the $\famC_2$ family and demonstrates that this is a stable librational family. The dashed circle indicates the
value of $e_1$ expected for these initial conditions. The red curve shows a direct integration of a trajectory
near this fixed point. The reason that the available parameter space is restricted to between the two magenta
circles is that the angular momentum restricts the allowed range of e$_1$ (as can be seen from Figure~\ref{fig:himufig}). In particular, the range is limited by the requirement $\theta>-1$, which means that the saddle point $\famO_{||}$ also appears. The dashed lines show the expected value of $\omega_1$ calculated from equation~(\ref{eqn:quadcop}). The blue curve shows an integration that starts from a coplanar, retrograde configuration for these parameters. We see that this saddle point is sensitive to the octupolar terms, as the orbit switches between libration and circulation.
}
\label{fig:C2D2}
\end{figure}

 Figure~\ref{fig:C2D2} shows the orbital behaviour near the $\famC_2$ and $\famD_2$ families of retrograde
orbits in the bottom right of Figure~\ref{fig:himufig}. We see that these correspond to a fixed point, i.e.
a stable equilibrium. This is therefore the high $e_2$ equivalent of $\famC_{\rm Q}^-$, just as $\famC_1$ and $\famD_1$ are the high
$e_2$ equivalents of the $\famC_{\rm Q}^+$ family.

This new set of families first appears at $e_1=0$ when the $\famC_{\rm Q}^-$ family passes through $e_1=0$ and $\theta=-0.856$.
This is also the point at which the special family $\famO_{\rm C}$ appears, with $\cos 2 \omega_1=-1$, and so this becomes
degenerate with the $\famC_{\rm Q}^-$ family. We see no equivalent split on the prograde side because $\cos 2 \omega_1=1$ would
require a much larger $\mu_{12}$ and the $\famC_{\rm Q}^+$ family does not pass through any such point.
A more general criterion for the appearance of this family can be obtained by noting that the $\famC_{\rm Q}^{-}$ family represents
the $e_2 \rightarrow 0$ limit of the $\famC_2$/$\famD_2$ families. 

	
\subsubsection{The Saddle Points: $\famA_1$, $\famB_{\rm Q}$, $\famB_1$, $\famC_3$ and $\famD_3$}

Also apparent in Figure~B7 is that there is a saddle point close to the origin -- these are the $\famA_1$ and $\famD_3$ stationary point families that are also present at lower masses.
 The families $\famB_1$ and $\famC_3$ were also present at lower masses, but were found at retrograde inclinations and moderate $e_1$. They have now shifted to an inclined family or almost circular orbits, and 
 are now saddle points. In the case of 
$\famC_3$ this is a change in the behaviour of this family, relative to the discussion in \S~\ref{sec:C3}.

\begin{figure}
                \centering
                \includegraphics[width=1.0\linewidth]{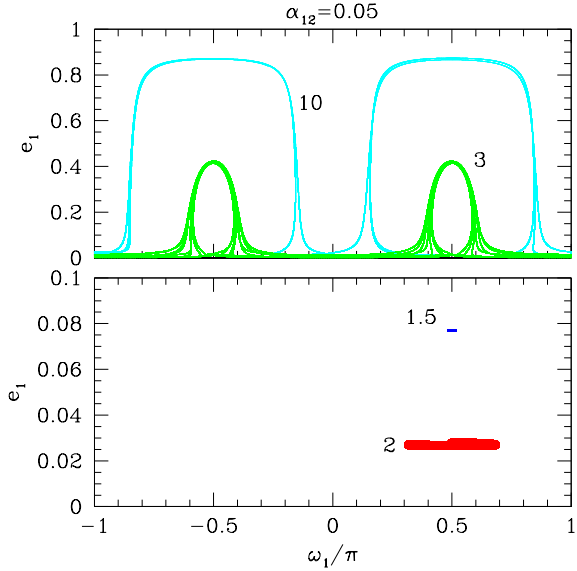}
                \caption{The lower panel shows the stable libration of $\omega_1$ for the cases of $\mu_{12}=1.5$ and 2. The upper panel shows how these stationary points become saddle points at higher masses, leading to large variations in $e_1$. The larger amplitude of libration of the red curve in the lower panel is the harbinger of looming instability, as it is driven by the larger amplitude libration of $\omega_2$, which eventually overwhelms $\omega_1$.
}
\label{fig:C3multi}
\end{figure}

 Figure~\ref{fig:C3multi} shows the evolution of e$_1$ versus $\omega_1$ in examples from the $\famC_3$ family for four different mass ratios -- $\mu_{12}=1.5$ from the last section,
$\mu_{12}=10$ from this section, and two intermediate values $\mu_{12}=2$ and $\mu_{12}=3$. We see that the transition
occurs between $\mu_{12}=2$ and 3, and results when the $\famC_3$ family switches from a retrograde family at small $e_1$, to a circular family slightly inclined relative to retrograde coplanar. Thus, increasing $\mu_{12}$ destabilizes the $\famC_3$ family, turning it from a fixed point to a saddle point. This is a consequence of the increased libration and eventually circulation of $\omega_2$, which causes $\omega_1$ to circulate, with only intermittent libration.

 The $\famB_{\rm Q}$ family also retains its saddle point nature, as expected (since $e_2$ is small along this family). This is shown in Figure~B8 of the online appendix, which shows both the $\famB_{\rm Q}$ point at $e_1 \sim 0.37$, but also the $\famC_{\rm Q}^+$ fixed point at $e_1 \sim 0.95$. Unlike the prior contour plots, this one does not allow solutions for the full range of $e_1$, because the dotted curve corresponding to the fixed angular momentum does not extend to $e_1=0$ in Figure~\ref{fig:himufig}.


\subsection{Extreme Mass ratios: $\mu_{12}=20$ and $\mu_{12}=100$}
\label{sec:ExMass}

Figure~\ref{fig:FamEvol} shows the evolution of the stationary point families as $\mu_{12}$ continues to get
larger and starts to approach the outer test particle limit. The behaviour of the stationary point families can be divided into several subsets.

\subsubsection{The Kozai-Lidov Generations: $\famC_{\rm Q}$, $\famC_1$, $\famC_2$; $\famD_1$ and $\famD_2$}
As shown in Figure~\ref{fig:FamEvol}, at $\mu_{12}=20$, the prograde families $\famC_1$ and $\famC_{\rm Q}^+$ are compressed to $e_1<0.2$ and they disappear
completely by $\mu_{12}=100$ (although vestigial versions of $\famC_{\rm Q}^-$, $\famC_2$ and $\famD_2$ remain). The
$\famD_1$ family comes to dominate at these masses, and becomes progressively more polar as the mass ratio
increases.

\subsubsection{The Saddle Points: $\famA_2$, $\famB_{\rm Q}$, $\famB_1$ and $\famD_3$}

 Similarly to $\famD_1$, the $\famB_{\rm Q}$ family evolves towards the polar limit as $\mu_{12}$ increases, but from the retrograde direction. 
       \begin{figure}
                \centering
                \includegraphics[width=1.0\linewidth]{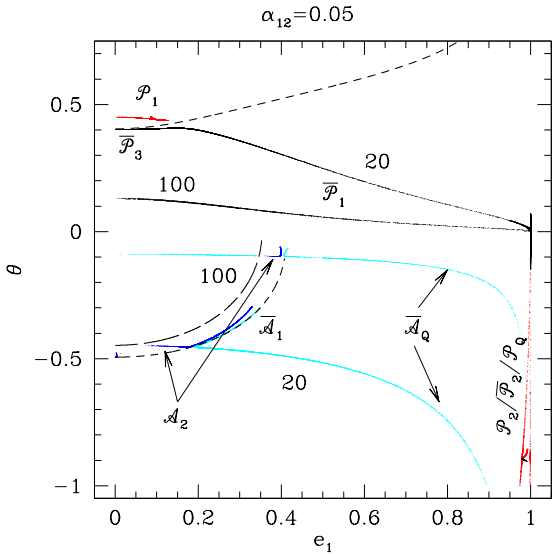}
                \caption{This plot shows the fixed point families for two cases -- $\mu_{12}=20$ and $\mu_{12}=100$. The colours represent the same apsidal configurations as before. The overall trend is to drive the families towards polar orbits as $\mu_{12}$ increases. The short dashed curves represent the $e_1$--$\theta$ relationships which cause the quadrupolar part of $\dot{\omega_2}$ to vanish. The long dashed curve is when the octupolar contribution to $\dot{\omega}_2$ vanishes in the $e_2 \rightarrow 0$ limit.
}
\label{fig:FamEvol}
\end{figure}
At larger $\mu_{12}$, the family $\famB_1$ moves to larger $e_1$ and truncates when it intersects $\famB_{\rm Q}$. Furthermore, we
finally see the appearance of the $\famA_2$ family discussed in \S~\ref{sec:CaseA}.  We see that it is also clearly associated with the $\famB_{\rm Q}$/$\famB_1$ family, filling in a gap in the $\famB_{\rm Q}$ family at $\mu_{12}=100$. 
The appearance of the $\famA_2$ family produces a new saddle point. This has qualitative
similarities to the $\famB_{\rm Q}$ point, as it divides the parameter space into an inner and outer region of circulation, which
encircle a libration (about the $\famC_1$ fixed point). This is demonstrated in Figure~B8 of the online appendix.


The $\famD_1$ family merges smoothly into the saddle point family $\famD_3$, so one question,
 based on Figure~\ref{fig:FamEvol} is whether, in the limit of large $\mu_{12}$, the prograde solutions are entirely of family $\famD_1$, or whether a family of $\famD_3$ saddle points remains. Examination of the black curve for $\mu_{12}=100$ shows that this entire curve is stable -- i.e. belongs to the $\famD_1$ family. Thus, $\famD_3$ disappears at the same time as $
\famC_1$.

\subsection{Influence of Relativity}

In \S~\ref{sec:Relative} we showed that the inclusion of relativistic precession alters the positions of the stationary point solutions if the inner planet orbits too close to the star. It also has consequences for the stability of those
equilibria.

\begin{figure}
                \centering
                \includegraphics[width=1.0\linewidth]{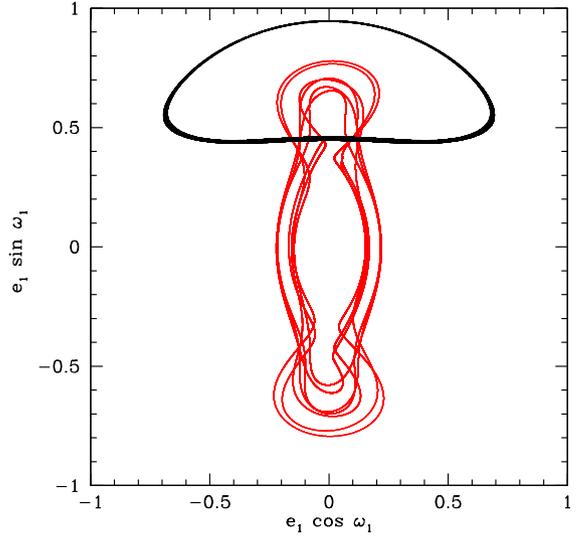}
                \caption{The red curve shows the result of a direct integration (including relativity) starting from the $\famC_2$ family in the middle panel of Figure~\ref{fig:3Pan}. In particular, $e_1=0.461$, $e_2=0.3965$ and $\theta=-0.633$ (with $\omega_1=\omega_2=\pi/2$). We see that this alternates between periods of libration and circulation, which are driven by the circulation of $\omega_2$ and the resulting fluctuations in $e_2$. The black curve shows the libration that results purely from the secular dynamics (no relativistic precession) using the exact same initial conditions.
}
                \label{fig:RelStab}
        \end{figure}

In particular, the inclusion of relativistic precession appears to destabilise both the quadrupolar extensions
$\famC_{\rm Q}$ as well as the $\famC_2$ and $\famD_2$ families seen in the panels of Figure~\ref{fig:3Pan}. An example of this is shown in Figure~\ref{fig:RelStab}. We see that $\omega_1$ does librate intermittently, but also experiences circulation and fluctuations in $e_1$, which are driven by the circulation of the angle $\omega_2$. Thus, we cannot regard this family as a fixed point family anymore.

The family that does remain stable is $\famC_1$, in which both $\omega_1$ and $\omega_2$ librate for starting conditions taken from all three panels in Figure~\ref{fig:3Pan}. The corresponding family $\famD_1$ in the lower panel shows libration
of $\omega_1$ but circulation of $\omega_2$, which introduces a larger amount of variation in $e_1$. 

This prograde, polar fixed point family is stable even when the quadrupolar family is destroyed by the relativistic precession, and extends down to at least $a_1=0.025$AU in the case where $M_1=M_2=1 M_J$ and $\alpha_{12}=0.05$. Of perhaps greater relevance is the case where the inner mass and semi-major axis is fixed and the outer values varied, as this is more representative of the observational situation. The sequence shown in Figure~\ref{fig:Rel3} is more representative of this, and shows that the $\famC_1$ family persists in this sequence as well. It also remains stable as $a_2$ and/or $M_2$ increases. However, it does move to ever higher $e_2$, although $\epsilon$ remains below the threshold level of 0.1 because
$\alpha_{12}$ is also dropping.

\subsection{Influence of Orbital Separation}

We have so far focussed on the mass ratio, $\mu_{12}$, as the principal parameter, holding the ratio
of separations fixed at $\alpha_{12}=0.05$. At the quadrupolar level, the stationary point structure is regulated by the quantity $\mu_{12} \alpha^{1/2}_{12}$, so that the configurations for smaller $\alpha_{12}$ should largely mimic those at $\alpha_{12}=0.05$ but with lower $\mu_{12}$. At the octupole level, an additional consideration is the fact that changing $\alpha_{12}$ will change the value of
$e_2$ corresponding to the threshold $\epsilon=0.1$. A comparison at fixed $\mu_{12}$ but different $\alpha_{12}$ is shown
in Figure~B11 of the online appendix. As expected, shifting from $\alpha_{12}=0.05$ to $\alpha_{12}=0.005$ moves the
families towards a configuration more reminiscent of the lower mass case in Figure~B5. Similarly, moving to
larger $\alpha_{12}=0.2$ shifts the configuration more towards that observed in Figure~\ref{fig:FamEvol}.




 If $\alpha_{12}$ gets too large, the neglect of higher order terms becomes problematic and the hierarchical assumption fails.
 However, at least some of the stationary points discussed here appear to survive at closer separations. Studies of the secular structure of specific exoplanet pairs by direct numerical averaging of the secular hamiltonian \citep{MFB06,MG09} show fixed points associated with Kozai-Lidov resonances as well as several additional families. Section~\S~B4 of the online appendix discusses the relationship between the naming convention used here and that used in \cite{MG09,MG11}. These
studies focus either on a limited mass range and more compact configurations \citep{MG09} or high masses but only prograde orbits \citep{MG11}, but indicate that the structure we discuss here is robust beyond the hierarachical approximation.

	\section{Discussion}
 \label{sec:discussion}

Table~\ref{StableTable} summarises the different stationary point families and clarifies which
are fixed points and which are saddle points. In terms of stable stationary points (fixed points), we find that the octupole problem shows
analogues of the quadrupolar Kozai-Lidov family ($\famC_{\rm Q}$, $\famD_{\rm Q}$) for low but finite $e_2$ as well as branches that exist at large $e_2$ ($\famC_1$ and $\famD_1$). We also find that these branches switch apsidal alignments in certain places, driven by the direction of the precession of $\omega_2$ at quadrupole order. We find a branch of fixed points at almost radial orbits ($\famC_{\rm R}$ and $\famD_{\rm R}$) and another stable family of fixed points in a retrograde configuration ($\famC_2$ and $\famD_2$). The stationary point structure is much richer
for $\mu_{12}>1$ than it is for $\mu_{12}<1$. 

An analogue of the $Q_1$ saddle point also appears at large $\mu_{12}$, in the form of $\famB_{\rm Q}$, $\famB_1$ and $\famA_2$, with branches for both small and large $e_2$ in a manner similar to the $\famC$/$\famD$ family fixed points. The set of saddle point families $\famA_1$, $\famB_1$, $\famC_3$ and $\famD_3$ define the seperatrices of large scale librations about the $\famC$ and $\famD$ fixed point families in the limit 
$e_1 \rightarrow 0$.
 There are also 
 generalisations 
of the special case saddle points for circular, radial and coplanar orbits.
	
	\subsection{Switching of Apsidal alignments}
\label{sec:switch}	
One curious feature of the solutions to the octupole problem is that, although extensions of the quadrupolar families are present, the generalised families switch apsidal alignment for
particular values of $e_1$.
 This can be understood by noting that the sign of the octupolar contribution must change whenever the sign of the quadrupolar contribution switches sign, in order to fix the stationary point. Thus, we can identify the switches in apsidal alignment by finding the cases when $\dot{\omega}_1=\dot{\omega}_2=0$ at quadrupole order.

At low $\mu_{12}$ there are no switches in apsidal alignment. As we increase $\mu_{12}$, the first case arises in the limit $e_1 \sim 1$. If
we take the $e_1 \rightarrow 1$ limits of the prograde versions of equations~(\ref{eqn:Quad2}) and (\ref{eqn:Quadouter}), we find the critical value at which this apsidal switch first appears from 
the requirement that the two expressions have the same limit, namely
\begin{equation}
    \frac{\mu_{12} \alpha_{12}^{1/2}}{2} \sqrt{\frac{1-e_1^2}{1-e_2^2}} \left( 1 + 
    \left( 1 + \frac{12 \sqrt{1 - e_2^2}}{5 \mu_{12}^2 \alpha_{12}} \right)^{1/2}\right) =
    \mu_{12} \alpha_{12}^{1/2} \sqrt{\frac{1-e_1^2}{1-e_2^2}}.
\end{equation}
This is satisfied when
  $\mu_{12}^2 \alpha_{12}=0.3$. Therefore, the two curves first overlap when $\mu_{12}>\sqrt{6}=2.449$ for $\alpha_{12}=0.05$. For mass ratios above this value, the prograde analogue of the $Q_2$ solution is split
 between $\famC_{\rm Q}$$^+$ and $\famD_{\rm Q}$$^+$, with the transition moving to lower $e_1$ as $\mu_{12}$ increases. 

As the mass ratio increases, the point of reversal moves to lower $e_1$ and eventually disappears again. The corresponding mass can be derived from the $e_1 \rightarrow 0$ limit, which yields $\mu_{12}^2 \alpha_{12} = 256/5 
(1-e_2^2)$, or $\mu_{12}>32 (1- e_2^2)$ for $\alpha_{12}=0.05$.

\begin{figure}
                \centering
                \includegraphics[width=1.0\linewidth]{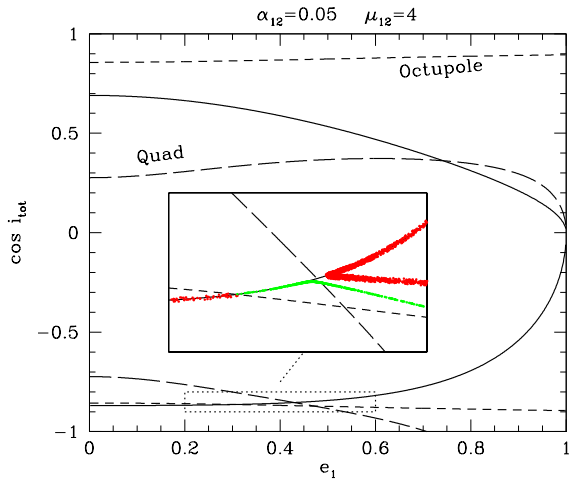}
                \caption{The solid line shows the criterion for $\dot{\omega}_1=0$ at the quadrupolar
                level (the $Q_2$ solution). The long dashed lines shows the equivalent criterion for
                $\dot{\omega}_2=0$. The short dashed lines show the criterion for the octupolar contribution
                to $\dot{\omega}_2=0$ in the limit of small $e_2$. The inset shows a zoom in to the region where
                the three curves cross in the retrograde case. In red we show the full $\famC$ family solutions for this
                region and in green we show the $\famD$ family. We see there is a limited range of $e_1$ for which there is an apsidal reversal -- sandwiched between the criteria for the quadrupolar and octupolar contributions to reverse. 
}
                \label{fig:quadzoom}
        \end{figure}

Similar behaviour occurs for the retrograde branches. In this case, the alignment reversal sets in at lower masses in the $e_1 \rightarrow 0$ limit. There is no
closed form solution for this criterion, but numerically it is $\mu_{12} \alpha_{12}^{1/2} = 0.659 \sqrt{1-e_2^2}$. As the mass increases, the reversal moves to larger $e_1$, with a limiting behaviour in the large $\mu_{12}$ limit of \begin{equation}
    e_1^2 \rightarrow 1 - \frac{2}{5} \frac{(1-e_2^2)}{\alpha_{12} \mu_{12}^2}. 
\end{equation}
The retrograde solutions exhibit another curious feature, in that the octupolar contribution to $\dot{\omega}_2$
 contains a 
term $\propto 1/e_2$. The $\famC_{\rm Q}$$^-$ and $\famD_{\rm Q}$$^-$ solutions occur in the limit of low $e_2$ so fixing a stationary point in this limit requires that the coefficient of that term must go to zero too, which imposes a condition
$\theta^2 = \frac{1}{5}(11+17 e_1^2)/(3+ 4 e_1^2)$ as well. When the solutions cross this threshold, there is another apsidal reversal. The consequence is that the $\famD_{\rm Q}$$^-$ contribution is limited to a finite range of $\mu_{12}$
and $e_1$. The inset in Figure~\ref{fig:quadzoom} illustrates this.

Related behaviour is apparent in the $\omega_1=0$ solutions. At the quadrupolar level, $\dot{\omega}_2=0$
yields equation~(\ref{eqn:Quadout2}).
Reversals in apsidal alignment are therefore to be expected when this criterion overlaps with the $Q_1$ criterion,
which leads to the condition
\begin{equation}
    \mu_{12}^2 \alpha_{12} = 16 \frac{1 - e_2^2}{1- 6 e_1^2}.
\end{equation}
The lowest mass for which a solution occurs is found by setting $e_1=0$, and so we expect apsidal reversals when $\mu_{12} > 4 \sqrt{(1-e_2^2)/\alpha_{12}}$,
which amounts to $\mu_{12}>17.9 \sqrt{1-e_2^2}$ for $\alpha_{12}=0.05$. We note also that solutions are 
limited to $e_1<1/\sqrt{6}=0.4082$ because otherwise $\mu^2_{12}<0$. The approach to this limit
leads to large $\mu_{12}$, which explains why this eccentricity was identified as a critical value
for the outer test particle case \citep{NLZ17,Zanardi+17,Zanardi+18,VC18,DeL18,Naoz+20}.

As in the case of the $\famC$ and $\famD$ families, we see that the apsidal reversal between $\famA$ and $\famB$ families also
only occupies a limited range of $e_1$, and for the same reason. The criterion that the octupolar term be finite
as $e_2 \rightarrow 0$ for this case implies a condition $\theta^2 = \frac{1}{5} (1 - 8 e_1^2)/(1-e_1^2)$. In
the limit of very large masses, $\theta \rightarrow 0$ and $e_1 \rightarrow 1/\sqrt{8} = 0.3536$. Together these
eccentricity limits explain the behaviour of the retrograde families in Figure~\ref{fig:FamEvol}.


\subsection{The Eccentric Kozai-Lidov Mechanism}

One motivation for this work was to get a more unified view of the rich dynamical structure \citep[e.g.][]{Naoz16} of the hierarchical
three body problem, using the stationary point families as a `scaffolding', or organising principle.
The most obvious feature of the dynamics of this problem is the Kozai-Lidov resonance \citep{Kozai62,Lidov62}, which couples the eccentricity and inclination variations due to a resonance between the apsidal and nodal precession rates. The original (quadrupole) description of this resonance is associated with our $Q_2$ family of fixed points, although the $Q_{\rm R}$ saddle point is also important, as it represents the turning point of the librations about the $Q_2$ family.


The appreciation that the octupolar contribution can qualitatively change the dynamics \citep{Naoz11,NF13} motivates the generalisation of the study of the stationary points to the more general octupolar case. We see that the form of the K-L fixed point family retains its basic nature with the introduction of the octupole term, although we note that the apsidal alignment between inner and outer orbits does vary depending on the eccentricity and mass ratio.

\begin{figure}
                \centering
                \includegraphics[width=1.0\linewidth]{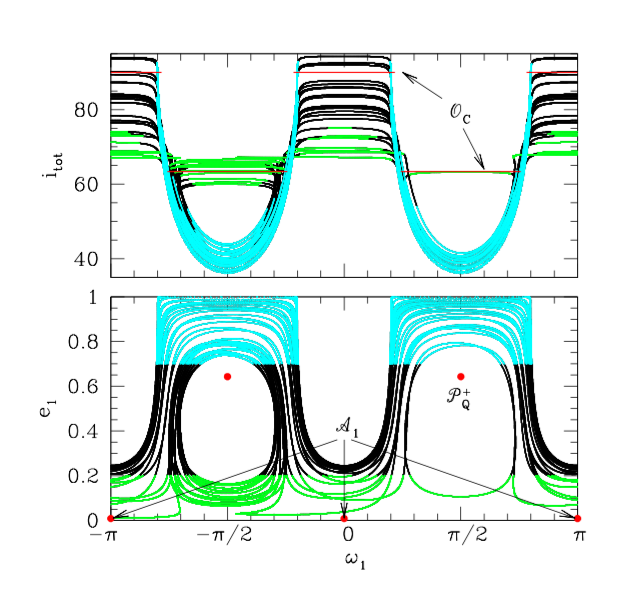}
                \caption{The upper panel shows the evolution of the mutual inclination, and the lower panel shows the evolution of the inner planet eccentricity e$_1$, as a function of the inner argument of perihelion. The parameters of the integration were chosen from Figure~6 of \citet{NF13}-- $m_c=1 M_{\odot}$, m$_1=1 M_J$, m$_2=2 M_J$, a$_1=4$AU, a$_2=45$AU, e$_1=0.01$, e$_2=0.6$, $\omega_1=\pi$ and $\omega_2=0$. Initial $i_{\rm tot}=67^{\circ}$. The green points indicate when $e_1<0.2$ and the cyan parts when $e_1>0.7$. The red points and lines indicate stationary point families, as labelled. In principle, the value of $e_1$ for $\famC_{\rm Q}^+$ should should vary with $e_2$, but the effect is small given the variation
observed here ($e_2$ varies from 0.496 to 0.618).
}
                \label{fig:Naoz13Fig6}
        \end{figure}

Figure~\ref{fig:Naoz13Fig6} shows the evolution of a system with the same parameters as those shown in Figure~6 of \cite{NF13} -- a demonstration of the kind of orbit orientation flip introduced by the inclusion of the octupole term in the dynamics. The evolution of the system is plotted relative to the particular stationary points relevant to the dynamics of this particular case. We see that the fundamental libration is still driven by the location of the $\famC_{\rm Q}^+$  family, but the turnaround at low e$_1$ is associated with the $\famA_1$ saddle point. It is also notable that the range of $\omega_1$ over which the value of e$_1$ remains low is regulated by two of the solutions of the $\famO_{\rm C}$ special family -- depending on the value of the mutual inclination. The excursions at large e$_1$ do come close to the $\famO_{\rm R}$ special point, but are regulated by the limited variation allowed by the conservation of angular momentum (even including the variations induced by libration of e$_2$).

The generalisation of the K-L family -- $\famC_{\rm Q}$ -- occurs for small, but non-zero $e_2$. We do find, however, that there are extensions of this fixed point family to high $e_2$ (the $\famC_1$/$\famD_1$ and $\famC_2$/$\famD_2$) families. In particular, the $\famC_1$ and $\famC_{\rm Q}^+$ families share very similar trends in terms of e$_1$ and $\theta$, but differ dramatically in terms of e$_2$. Are they truly distinct? 

\begin{figure}
                \centering
                \includegraphics[width=1.0\linewidth]{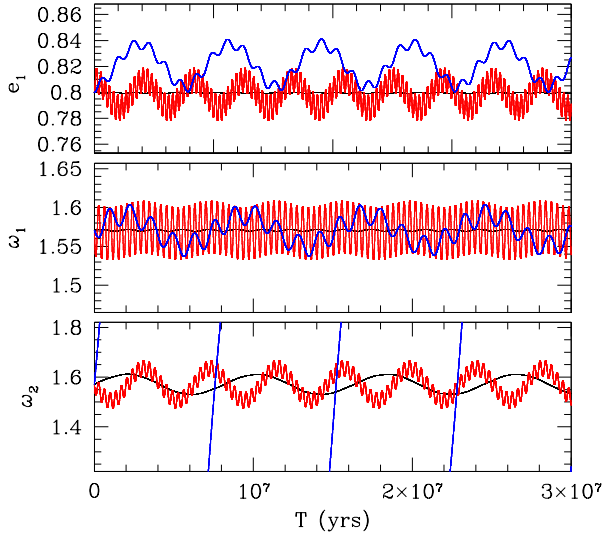}
                \caption{The upper panel shows the evolution of $e_1$ as a function of time for each of the three cases
discussed in the text.
 The middle panel shows that the inner argument of periastron librates for all cases studied here, while the lower panel shows the outer argument of periastron. We see that this angle librates for  $e_2=0.03$ (the $\famC_Q$$^+$ case -- black curves) and $e_2=0.77$ (the $\famC_1$ case -- red curves) but circulates for $e_2=0.4$ (blue curves). 
}                \label{fig:EKL}        \end{figure}

 Figure~\ref{fig:EKL} shows the results of three integrations that all start from the same initial conditions except for a different value for e$_2$. The masses are $m_c=1 M_{\odot}$, $m_1=2 M_J$ and $m_2=1 M_J$, while the semi-major axes are $a_1=1$AU and $a_2=20$AU (so $\alpha_{12}=0.05$). The inner eccentricity is $e_1=0.8$ and $i_{\rm tot}=67.6^{\circ}$. We also assume $\omega_1=\omega_2=\pi/2$.
For these values, the $\famC_{\rm Q}^+$ fixed point is located at $e_2=0.03$ and the $\famC_1$ fixed point is located at $e_2=0.77$.
 The figure shows the evolution for e$_2$=0.03 (black), 0.4 (blue), and 0.77 (red).


For low values of e$_2$, $\omega_2$ librates, as expected for the $\famC_{\rm Q}$$^+$fixed point. However, for 
$e_2$ in the range 0.1--0.7, $\omega_2$ circulates, although $\omega_1$ continues to librate.
This is consistent with the behaviour expected of a generalization of the quadrupole behaviour, since $e_2$ plays little role in the evolution of $\omega_1$ if the octupolar contribution is weak. However, Figure~\ref{fig:EKL} shows
that libration of $\omega_2$ returns at $e_2=0.77$. This is the appearance of the $\famC_1$ fixed point.
Thus, the $\famC_1$ point is qualitatively distinct from the $\famC_{\rm Q}$ point in the sense that we find an extended
range of e$_2$ in between the two fixed point values, for which  $\omega_2$ circulates.

\subsection{Coplanar Flip Behaviour}
\label{sec:CopFlip}

\cite{LN14} noted the appearance of an  orbital `flip' behaviour in systems with almost coplanar orbits
but high eccentricities. This is qualitatively different from that associated with the Kozai-Lidov family in
that it starts from approximately coplanar configurations and also transitions from prograde to retrograde on
a timescale considerably shorter than the diffusive evolution seen in manifestations of the Eccentric Kozai-Lidov effect.

This appears to be related to the radial fixed point family $\famO_{\rm R}$ discussed in \S~\ref{sec:OctoSpecial}. Li et al discussed the case of $\alpha_{12}=0.02$ and find that the orbit flips if the outer body eccentricity is large enough.
 The family $\famO_{\rm R}$ is a saddle point, and so the approach to this limit in the coplanar case drives the system away from the equilibrium. As noted in \S~\ref{sec:ORlimit}, the range of $\omega_1$ for which the equilibrium exists is limited unless e$_2$ is large enough.
If we apply our criterion equation~(\ref{eqn:ORcrit}) for the critical solution to extend over all $\omega_1$, we derive
a criterion $e_2>0.627$, which compares well to Li et al's empirical estimates for the threshold value. They also
find a restriction to large initial $e_1$, but this is more related to the initial conditions -- it is required to limit the value of the z-component of the angular momentum to a value that is less than the variation induced by the octupolar term. 

   \begin{figure}
                \centering
                \includegraphics[width=1.0\linewidth]{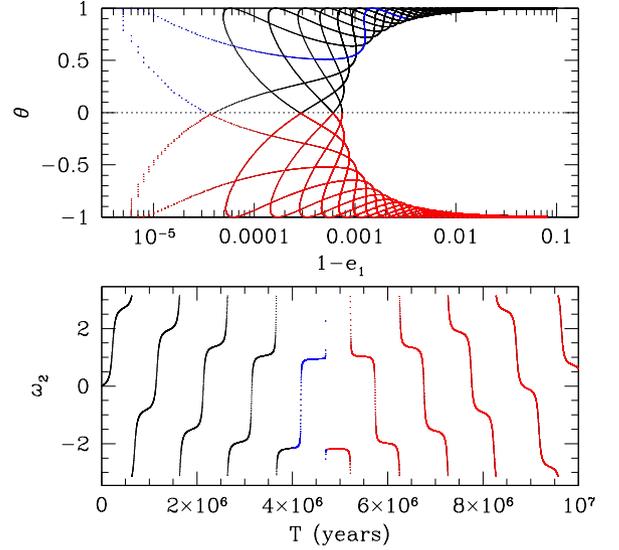}
                \caption{The upper panel shows an example of the `Coplanar flip' discussed in \citet{LN14}. The parameters here are chosen to reproduce Figure~2 of that paper. The masses are $M_c=1 M_{\odot}$, $M_1=0.001 M_{\odot}$, $M_2=0.02 M_{\odot}$ and the semi-major axes are $a_1=1$AU and $a_2=50$AU. The initial eccentricities are $e_1=0.9$ and $e_2=0.7$. The initial mutual inclination is $5^{\circ}$ and the arguments of periastron are chosen to be $\omega_1=\omega_2=0$ in the
invariable plane. The lower panel shows the precession of the angle $\omega_2$ in the neighbourhood of the first flip. The curves are red if $i_{\rm tot}>90^{\circ}$. We see that the flip is associated with a reversal in the direction of precession of $\omega_2$ i.e., it has passed through the saddle point at $\theta=0$ and $e_1 \sim 1$. The last cycle in $\omega_2$ is highlighted in blue and demonstrates that the transition from almost coplanar to flip is rapid -- as noted by Li et al., who found this happened much more rapidly than in the Kozai-Lidov case.
}                \label{fig:CopFlip}        \end{figure}

Figure~\ref{fig:CopFlip} shows an integration of the case chosen by \cite{LN14} in Figure~2 of that paper. We see that the `flip' is indeed associated with a saddle point in $\omega_2$ -- where the direction of precession reverses. For lower values of e$_2$, the precession of $\omega_2$ does not reverse and the inclination increases by only a small amount.
We also highlight the  cycle in $\omega_2$ that leads to the flip in blue. This demonstrates the speed at which the transition occurs -- because it is a direct passage through the saddle point and not a diffusive evolution like in the case of the flips driven by the Eccentric Kozai-Lidov evolition.


\subsection{The Inverse Kozai-Lidov Resonance}

\begin{figure}
                \centering
                \includegraphics[width=1.0\linewidth]{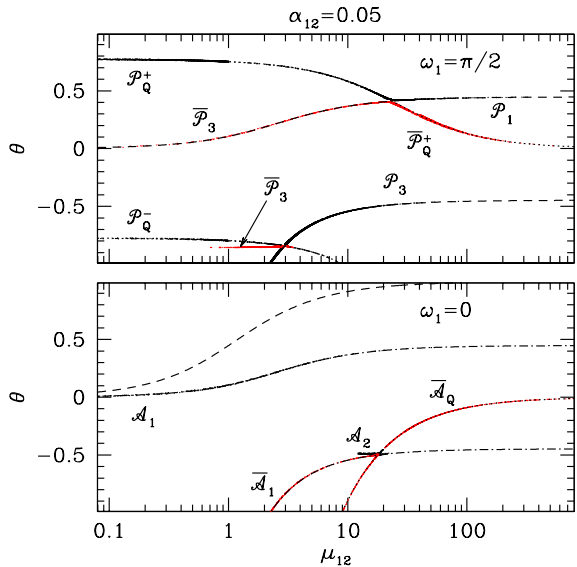}
                \caption{The upper panel shows the critical inclinations ($e_1=0$) for the orbital families in the
case $\omega_1=\pi/2$. Black points are the apsidally aligned families, and red points are the apsidally anti-aligned.
The lower panel shows the equivalent for the $\omega_1=0$ cases. The dotted lines indicate the 
quadrupole constraint on $\dot{\omega}_1=0$, the dashed curve incidates the quadrupolar term $\dot{\omega}_2=0$
and the dot-dashed curve indicates the octupole term $\dot{\omega}_2=0$.}
                \label{fig:MuyC}        \end{figure}

 The original Kozai-Lidov solution represents the fixed point associated with an inner test particle, and is found above a
critical inclination (at low $e_1$) of $\theta^2=3/5$. The equivalent solution for an outer test particle yields another 
 critical inclination at $\theta^2=1/5$ \citep{JM66,Kras72,LZ76} at quadrupole order. Figure~\ref{fig:MuyC} shows the critical inclinations $I_{\rm crit}$ (the $e_1 \rightarrow 0$ limit) of various orbital families, as a function of $\mu_{12}$ (keeping $\alpha_{12}=0.05$).
The changeover from the $\famC_{\rm Q}$ family to the $\famC_1$ contribution is because
 $\dot{\omega}_1=0$ and $\dot{\omega}_2=0$ have different quadrupolar limits as the
mass ratio increases, and it requires large octupolar corrections to satisfy both criteria simultaneously. Eventually
this violates the $\epsilon<0.1$ criterion, although the addition of higher order terms can recover a solution in
this limit \citep{GGP12,NLZ17,VC18,DeL18}.

Within context of our classification, the `inverse Kozai-Lidov resonance' (as defined by \cite{VC18}) is related
to the high-$e_2$ extensions of the original quadrupolar family ($\famC_1$ for the prograde case, $\famC_3$ for
the retrograde case) rather than the low $e_2$ extensions $\famC_{\rm Q}^+$ and $\famC_{\rm Q}^-$ that form the
natural generalisations of the Kozai-Lidov family, as these latter families either tend to polar or disappear in
the high $\mu_{12}$ limit.

\subsection{Warm Jupiters}

The presence of giant planets on scales $\sim 0.1$--1~AU (`Warm' Jupiters) is considered to be a curious phenomenon, given that the
most common theories of giant planet formation suggest that planets are easier to form on larger scales. The presence
of `Hot' Jupiters, with $a<0.1$AU is suggested to be a consequence of either migration through a disk to the inner
edge \citep{GT80, LP86, LBR96} or by the tidal capture of planets excited to high eccentricity orbits due to either planetary
scattering \citep{RF96,WM96} or secular interactions \citep[e.g.,][]{FT07,Naoz11,WL11,Naoz+12,Stephan+18}. Warm Jupiters fall in between these two classes. It has been proposed that such
planets may be in the process of a slow or stalled tidal drag-down because their periastra get close enough to the star
for meaningful tidal dissipation only for a small fraction of the duration of secular oscillations \citep{DKS14, PT16, FH16}. 
\cite{DC14} present evidence that Warm Jupiters with outer planetary companions have substantial mutual inclinations,
based on the clustering of the projected $\delta \omega$ near values $\sim 90^{\circ}$. This clustering is related to
the fact that systems undergoing large librations about the $\famC_{\rm Q}^+$ family approach the $\famO_{\rm C}$ saddle point and the
preferred value of $\theta \sim 0.77$ emerges from the location of the $\famC_{\rm Q}^+$ family in the limit of $\mu_{12}<1$
(which holds for most of the Warm Jupiter systems). 




\subsection{High Eccentricity Orbits}

Much of the interest in hierarchical triples derives from their potential to generate high eccentric orbits through Kozai-Lidov oscillations \citep{Naoz16}. One new feature identified here is the existence of highly eccentric fixed point families, where the eccentricity remains high. The $\famC_{\rm R}$ and $\famD_{\rm R}$ families exist for approximately polar orbits, while the $\famC_2$ and $\famD_2$ families exist for retrograde orbits in the limit of large $\mu_{12}$. These families of orbits potentially offer alternative pathways to high eccentricity migration, but would require a dissipative process to place a system into such a configuration if one started from a traditional coplanar alignment.

\section{Conclusions}
\label{sec:Conclude}

Our goal in this paper is to survey the stationary points of the hierarchical three body problem. We aim to
 understand the variety of possible secular behaviours available to planetary systems in hierarchical configurations.

The principal feature at the quadrupole level of approximation is the fixed point family $Q_2$, identified originally
by Lidov and Kozai \citep{Kozai62,Lidov62}, along with several saddle points that appear in various limits \citep{Zig75,LZ76}. 
We find that the same fixed point behaviour appears at the octupolar level, for small but non-zero values of the
outer planet eccentricity e$_2$, although it is split between  two fixed point families, $\famC_{\rm Q}$ and $\famD_{\rm Q}$, depending on whether $\omega_1$
and $\omega_2$ are aligned or anti-aligned. An interesting feature of these families is that, for larger mass ratios, the apsidal alignment can change as a function of $e_1$. These switches are associated with the change in sign of $\dot{\omega}_1$ and $\dot{\omega}_2$ at quadrupole order. 

In addition to identifying the analogue to the quadrupolar family, we also identify extensions to this family,
with both prograde and retrograde cases. These new fixed point families are distinct in that they are branches that
continue up to much larger $e_2$ and have no analogue at the quadrupolar level. They demonstrate that 
the octupolar contribution can do more than simply induce variations about the quadrupole solution
and contribute to chaos -- it can also help to fix and stabilise new equilibria. 
Elements of these families are also the most robust against the destabilising effects of relativistic precession when the
inner planet gets close to the central star. 

The secular architecture  gets more diverse as $\mu_{12}$ increases (at fixed $\alpha$), with the various new fixed points appearing for $\mu_{12}>1$ and the original Kozai-Lidov family tending to polar orbits as $\mu_{12}$ gets large.  We also find several stationary points for retrograde configurations.

We also find a variety of special case solutions, most of which are
saddle points (although we recover the known extension of the Laplace-Lagrange solution in the coplanar limit). As one would expect, many 
 of these are associated with the transitions between circulation and libration about one of the fixed points.
 One saddle point of dynamical significance is that in the radial, coplanar limit, which is responsible for the coplanar
flip behaviour identified by \cite{LN14}. 

These results indicate that the secular architecture of multiplanet systems contains several possible fixed points, especially in the case of more massive inner planets. Although current methods of planet detection yield only weak constraints on mutual inclination in most cases, the anticipated astrometric information to be gained from GAIA in the near future \citep{Perry14} may allow us to constrain the full three dimensional behaviour of the best studied systems and to classify their dynamics in terms of the behaviour outlined here.

This research has made use of NASA’s Astrophysics Data System and of the NASA Exoplanet Archive, which is operated by the California Institute of Technology, under contract with the National Aeronautics and Space Administration under the Exoplanet Exploration Program. S.N. acknowledge partial support from the NSF through grant No. AST- 1739160. Moreover, S.N. thanks Howard and Astrid Preston for their generous support. The authors acknowledge a helpful referee report. 

{\bf Data availability}: All data used in this paper are available upon request from the corresponding author.
	
	\bibliographystyle{mnras}
	\bibliography{refs}

\appendix

\section{Octupole Equations}
\label{eqn:octupole}

Our analysis of the stationary points is based on the equations in section~8 of \cite{Naoz16}. For completeness we
reproduce the equations here. Some simplifying quantities are 
\begin{eqnarray}
B &=& 2 + 5 e_1^2 - 7 e_1^2 \cos 2 \omega_1 \\
A &=& 4 + 3 e_1^2 - \frac{5}{2} B (1 - \theta^2) \\
\cos \phi & =& - \cos \omega_1 \cos \omega_2 - \theta \sin \omega_1 \sin \omega_2.
\end{eqnarray}

We can simplify our expressions a little because we are interested in the zeroes of the equations and so
 the absolute timescales is not of immediate relevance. We consequently 
divide out the quantity $6 C_2/G_1$ from \cite{Naoz16}. After this operation, the precession of the inner body
is given by 
\begin{equation}
\dot{\omega}_1 = \mathcal{F}_1 + \mathcal{F}_2 \cos 2 \omega_1 - \frac{5}{8} \frac{\alpha_{12} e_2}{(1-e_2^2)}
\left[ \mathcal{F}_3 + \mathcal{F}_4 \right]
\end{equation}
where
\begin{eqnarray}
\mathcal{F}_1 &=& 5 \theta^2 + e_1^2 - 1 +  \left( 2 + 3 e_1^2 \right) \beta \theta \\
\mathcal{F}_2 &=& 5 \left( 1 - \theta^2 - e_1^2 - \beta e_1^2 \theta^2 \right) \\
\frac{\mathcal{F}_3}{e_1 (\theta + \beta)} &=& \sin \omega_1 \sin \omega_2 \left[ 10 (3 \theta^2-1)(1-e_1^2) + 
  A\right] \nonumber \\
&& - 5 B \theta \cos \phi \\
\frac{e_1 \mathcal{F}_4}{(1-e_1^2)} &=& \sin \omega_1 \sin \omega_2\left[ 10 \theta (1-\theta^2)(1-3 e_1^2)\right] \nonumber \\
&& + (3A + 2 - 10\theta^2)\cos \phi
\end{eqnarray}
where $\beta = G_1/G_2 = \mu_{12} \alpha_{12}^{1/2} \sqrt{(1-e_1^2)/(1-e_2^2)}$. Note that we have assumed
$m_1,m_2 \ll m_0$ in calculating the prefactor of $\mathcal{F}_3$ and $\mathcal{F}_4$, so that this applies
primarily to the planet problem. 

The precession of the outer body is given by
\begin{equation}
\dot{\omega}_2 = \mathcal{F}_5 + \mathcal{F}_6 \cos 2 \omega_1 - \frac{5}{8} \frac{e_1}{(1-e_2^2)}
\left[ \mathcal{F}_7 + \mathcal{F}_8  \right]
\end{equation}
where
\begin{eqnarray}
\mathcal{F}_5 &=& (2+3e_1^2)\left[\frac{\beta}{2}(5 \theta^2-1) + \theta \right] \\
\mathcal{F}_6 &=& 5 e_1^2\left[ \frac{\beta}{2}(5 \theta^2-3) - \theta \right] \\
\mathcal{F}_7 &=& \sin \omega_1 \sin \omega_2 \left[ \frac{\beta}{e_2} (1+4 e_2^2) 10 \theta (1 - \theta^2) (1-e_1^2)\right. \nonumber \\
&& - \left. e_2 \left( 1 + \beta \theta \right) \left( A + 10 (3 \theta^2-1)(1-e_1^2) \right) \right] \\
\mathcal{F}_8 &=& \cos \phi \left( 5 B \theta e_2 (1+ \beta \theta) + \frac{(1+4 e_2^2)}{e_2} \beta A \right)
\end{eqnarray}

To complete the description of the stationary points, we also need the rate of change of the eccentricities,
\begin{equation}
\frac{\dot{e}_1}{1-e_1^2}=\mathcal{F}_9 \sin 2 \omega_1 - \frac{5}{8} \frac{\alpha_{12} e_2}{1-e_2^2}
 \left( \mathcal{F}_{10}  + \mathcal{F}_{11}  \right)
\end{equation}
where
\begin{eqnarray}
\mathcal{F}_9 = && 5 e_1 (1-\theta^2)  \\
\mathcal{F}_{10} =&& 35 \cos \phi (1-\theta^2) e_1^2 \sin 2\omega_1 \\
\mathcal{F}_{11} = && \theta \left[ A - 10 (1-\theta^2) (1-e_1^2) \right] \cos \omega_1 \sin \omega_2 \nonumber \\
&& - A \sin \omega_1 \cos \omega_2
\end{eqnarray}
and
\begin{equation}
\frac{\dot{e}_2}{1-e_2^2} =  \frac{5}{8} \frac{\alpha_{12} e_1}{1-e_2^2} \beta \mathcal{F}_{12}
\end{equation}
where
\begin{eqnarray}
\mathcal{F}_{12} & =& \theta \left( 10 (1-\theta^2) (1-e_1^2) -1 \right) \sin \omega_1 \cos \omega_2 \nonumber \\
&& + A \cos \omega_1 \sin \omega_2
\end{eqnarray}

\end{document}